\documentclass{elsart}

\usepackage[tbtags]{amsmath}
\usepackage{amscd}
\usepackage{amsfonts}
\usepackage{epsfig}

\renewcommand{\etal}{\emph{et al.}}

\renewcommand{\mod}{\;\mbox{mod}\;}
\renewcommand{\vec}[1]{\mbox{\boldmath $#1$}}
\newcommand{\mat}[1]{{\mbox{\bfseries \rmfamily#1}}}
\newcommand{\Idmat}{\vec{1}}
\newcommand{\diag}{{\rm diag}}
\newcommand{\transp}{{\rm t}}
\newcommand{\norm}[1]{\left\| {#1} \right\|}

\allowdisplaybreaks

\begin{document}

\begin{frontmatter}

\title{Dynamic Fitness Landscapes in Molecular Evolution}

\author{Claus O.\ Wilke}
\address{
Digital Life Laboratory\\
Mail Code 136-93, Caltech\\
Pasadena, CA 91125
}

\author{Christopher Ronnewinkel}\and
\author{Thomas Martinetz}

\address{
Institut f\"ur Neuro- und Bioinformatik\\
Medizinische Universit\"at L\"ubeck\\
Seelandstra\ss e 1a\\
D-23569 L\"ubeck, Germany
}

\begin{keyword}
dynamic fitness landscape, quasispecies, error threshold, molecular evolution\\
PACS: 87.23.Kg 
\end{keyword}

\date{\today}

\begin{abstract}
We study self-replicating molecules under externally varying
conditions.  Changing conditions such as temperature variations and/or
alterations in the environment's resource composition lead to both
non-constant replication and decay rates of the molecules. In general,
therefore, molecular evolution takes place in a dynamic rather than a
static fitness landscape. We incorporate dynamic replication and decay
rates into the standard quasispecies theory of molecular evolution,
and show that for periodic time-dependencies, a system of evolving
molecules enters a limit cycle for $t\rightarrow\infty$. For fast
periodic changes, we show that molecules adapt to the time-averaged
fitness landscape, whereas for slow changes they track the variations
in the landscape arbitrarily closely. We derive a general
approximation method that allows us to calculate the attractor of
time-periodic landscapes, and demonstrate using several examples that
the results of the approximation and the limiting cases of very slow
and very fast changes are in perfect agreement. We also discuss
landscapes with arbitrary time dependencies, and show that very fast
changes again lead to a system that adapts to the time-averaged
landscape. Finally, we analyze the dynamics of a finite population of
molecules in a dynamic landscape, and discuss its relation to the
infinite population limit.
\end{abstract}

\end{frontmatter}

\newpage
\tableofcontents
\newpage

\section{Introduction}
Eigen's quasispecies model~\cite{Eigen71} has been the basis of a
vivid branch of
molecular evolution theory ever since it has been put forward almost
30 years
ago~\cite{ThompsonMcBride74,JonesEnnsRangnekar76,Jones79a,Jones79b,%
EigenSchuster79,SwetinaSchuster82,McCaskill84,Rumschitzki87,%
Leuthaeusser87,Eigenetal88,SchusterSwetina88,Eigenetal89,NowakSchuster89,%
FranzPeliti97,AlvesFontanari98,WilkeRonnewinkelMartinetz99,NilssonSnoad2000,%
BaakeGabriel99}.
Its two main statements are the formation of a
quasispecies consisting of several molecular species with well defined
concentrations, and the existence of an error threshold above which
all information is lost because of accumulating erroneous mutations. Both
effects have been found in a number of experimental as
well as theoretical studies (For experimental evidence, see, e.g.,
\cite{Domingoetal78} for the formation of a
quasispecies in the RNA of the Q$\beta$ phage, \cite{AdamiBrown94} for
the observation of an error threshold in a system of self-replicating
computer programs, and \cite{Gomezetal99} for a more recent example
of quasispecies formation in the Hepatitis C virus. Generally, see the
reviews~\cite{Eigenetal88,Eigenetal89,BaakeGabriel99} and the references
therein). Recently, a new aspect of the quasispecies model has been
brought into consideration that was mostly missing in previous works,
namely the aspect of a dynamic fitness
landscape~\cite{WilkeRonnewinkelMartinetz99,NilssonSnoad2000,%
RonnewinkelWilkeMartinetz2000,NilssonSnoad2000a,NilssonSnoad2000b}
The notion ``Dynamic fitness landscape'' encompasses all situations in which
the replication and/or decay rates of the molecules change over time. In the
present work, we are interested in situations where these changes
occur as an external influence for the evolving system, without
feedback from the system to the dynamics of the
landscape. Dynamic fitness landscapes of this kind are important,
because almost any biological system is subject to external changes in
the form of, e.g., daytime/nighttime, seasons, long-term climatic changes,
geographic changes due to tectonic movements, to name just a few.

The main problem one encounters when dealing with dynamic landscapes
is the difficulty to find a good generalization of the quasispecies
concept. In the original work of Eigen, the quasispecies is the
equilibrium distribution of the different molecular species. It is reached
if the system is left undisturbed for a sufficiently long
time. Since in a dynamic landscape the system is being disturbed
by the landscape itself, the concept of a quasispecies is
meaningless in the general case. However, there are special
cases in which a meaningful quasispecies can be defined. If, for
example, the landscape changes on a much slower time scale than what the
system needs to reach the equilibrium, then the system is virtually in
equilibrium all the time, and the concentrations at time
$t$ are determined from the landscape present at that time.
Generally, it is certain symmetries in the dynamics of the
landscape that allow for the definition of a quasispecies. One example
we treat in this paper in detail is the case of time-periodic
landscapes, which offer a natural quasispecies definition.

An early investigation of dynamic landscapes has been carried out by
Jones~\cite{Jones79a,Jones79b}, who has only considered cases in
which all replication rates change by a common factor. Therefore, this
approach excludes (among other cases) in particular all situations in
which the order of the molecules' replication rates changes over time,
i.e., in which for example one of the faster replicating molecules becomes
one of the slower replicating molecules and vice versa.
Recent work on dynamic fitness landscapes allow for such
changes. Wilke \etal~\cite{Wilke99,WilkeRonnewinkelMartinetz99} have
developed a framework that allows to define and to calculate
numerically a quasispecies in time-periodic landscapes. Independently, Nilsson
and Snoad~\cite{NilssonSnoad2000} have studied the
particular example of a stochastically jumping peak in an otherwise
flat landscape. This work has been generalized by
Ronnewinkel \etal~\cite{RonnewinkelWilkeMartinetz2000}, who could also define a
meaningful quasispecies for a deterministic version of the jumping
peak landscape and related landscapes. Very recent work of Nilsson and Snoad
deals with self-adaptation of the mutation rate in a jumping peak
landscape~\cite{NilssonSnoad2000b}, and with the low-pass filter effect of
dynamic fitness landscapes~\cite{NilssonSnoad2000a}, a notion put forward by
Hirst~\cite{Hirst97a,HirstRowe99}. Finally, theoretical studies of
dynamic fitness landscapes can also be found in the related field
of genetic algorithms. Schmitt
\etal~\cite{Schmittetal98,SchmittNehaniv99} derive
results for finite populations in a relatively broad
class of dynamic landscapes. However, only
landscapes in which the fitnesses get scaled can be treated, so that the same
restriction applies here that applied to Jones's work. The order of the
fitnesses can never change. Also, genetic
algorithms with time-periodic landscapes have been studied by
Rowe~\cite{Rowe99a,Rowe99b}. However, Rowe's approach has
the disadvantage that it is tightly connected to the discrete time used in
genetic algorithms, and that the dimension of the transition matrices
grows in proportion to the period length $T$ of the oscillation. This
makes it hard to derive analytical results, and in addition to that,
it renders landscapes with large $T$ inaccessible to numerical calculations.

In this report, we do not cover the large field of molecular
evolution in the particular landscapes induced by RNA sequence-to-structure
mappings~\cite{Fontanaetal93,FontanaSchuster98,Forst98}, and, in connection to
that, we do not discuss neutral
networks~\cite{ForstReidysWeber95,HuynenStadlerFontana96,%
ReidysStadlerSchuster97,Nimwegenetal99}. We
do so mainly because these topics have so far not been studied in the light of
temporal variations in the fitness landscapes, and hence, a discussion of them
does not fit very well into the general tenor of this work. In general, it can
be argued that neutral networks are of less importance in a dynamic
environment, because in that situation a population is constantly on the move
to the next local optimum~\cite{WilkeMartinetz99a}.

The remainder of this report is structured as follows. We begin our
discussion in Section~\ref{sec:time-dependent-rep-rates} with a brief
summary of the general aspects of dynamic
fitness landscapes in the quasispecies model. In
Section~\ref{sec:periodic-fitness-landscapes}, we will develop the
main subject of this work: a general theory of time-periodic fitness
landscapes. The theoretical part thereof is presented in
Section~\ref{sec:diff-equ-formalism}, in which we demonstrate how a
time-dependent quasispecies can be defined by means of the monodromy
matrix, and how this monodromy matrix can be expanded in terms of the
oscillation period $T$. In Section~\ref{sec:discrete-approx}, we
present an alternative approximation formula for the monodromy matrix
that is more suitable for numerical calculations, and in
Section~\ref{sec:example-landscapes}, we compare, for several example
landscapes, the results obtained from that formula with the general
theory developed in Section~\ref{sec:diff-equ-formalism}. The
restriction of a time-periodic fitness landscape is weakened in
Section~\ref{sec:aperiodic-landscapes}, where we discuss the
implications of our findings for other, non-periodic fitness
landscapes. Since our work is based on Eigen's deterministic approach
with differential equations, all results presented up to the end of
Section~\ref{sec:aperiodic-landscapes} are only valid for infinite
population sizes. In order to address this shortcoming, in
Section~\ref{sec:finite-populations} we give a brief introduction into
the problems involved when dealing with finite populations. In
Section~\ref{sec:numerical-results}, some simulation results are
shown, demonstrating the relationship between the results
from the infinite population limit and the actual finite population
dynamics. Finally, an approximative analytical
description of a finite population evolving on a simple
periodic landscape is developed in
Section~\ref{sec:ana-finite-pop}. We close this paper with
some conclusions in Section~\ref{sec:conclusions}.

\section{The quasispecies model}

\subsection{Static landscapes}
With his model of
self-replicating molecules, Eigen showed for the first time that
Darwin's idea of mutation and selection can work in a simple, seemingly
``lifeless'' system of chemical reactants. His observation that evolution is
governed by the laws of physics spawned a fruitful field of work, and hundreds
of papers based on his initial ideas have been written since then. Most of
that work has been concerned with static fitness landscapes. There exist
comprehensive reviews of that work (see~\cite{Eigenetal88,Eigenetal89} for a
very detailed coverage of the
literature till 1989, and~\cite{BaakeGabriel99} for a more recent review).
Here, we are going to briefly introduce the main concepts developed for static
landscapes. In doing so, we restrict ourselves to those concepts that we will
refer to later on in our discussion of dynamic fitness landscapes. For a more
in-depth discussion of static landscapes, the reader is referred to the above
mentioned reviews.

The quasispecies model was originally aimed at describing self-replicating DNA
or RNA molecules. Therefore, the molecules were conceived of as sequences
consisting of a limited number of elementary building blocks. With that
picture in mind, we may represent the molecules as sequences of letters. For
RNA molecules, e.g., the conventional alphabet
consists of the 4 letters G, A, C, U, representing the 4 bases guanine,
adenine, cytosine, and
uracil, respectively. Today, most researchers are interested in the
information-theoretic aspects of the model. Consequently, the most common
alphabet in the molecular evolution literature has become
the binary alphabet, consisting of the letters 0 and 1. Throughout
this work, we will also adopt this choice. With regard to the RNA example,
the binary alphabet can be considered as distinguishing only
between purines (which are guanine and adenine) and pyrimidines
(which are cytosine and uracil).

The molecules the model describes have the ability to
self-replicate. Self-replication is a complicated process, which
consumes energy and substrates from the environment. These
external resources are supposed to be present, and are not
modeled explicitly. The degree to which a macromolecule finds the
resources necessary to self-replicate, and is able to
exploit them, is expressed in the replication coefficients
$A_i$. A molecule $i$ that finds good conditions for self-replication
has a high $A_i$, another molecule $j$ which is a bad self-replicator
has a much smaller $A_j$.

The molecules replicate by copying themselves. The copy procedure is
generally not error-free. Among the different types of errors one can
imagine for the copy of a sequence of letters (substitutions,
insertions, deletions), we consider only substitutions.
Substitutions can in principle occur with a different probability at
every single position in the sequence. However, if we assume the copy
procedure to be performed step by step by some sort of molecular
machinery, the probability of copying a symbol incorrectly can be expected
to be the same for all positions in the sequence. Hence, a letter will
change from 0 to 1 or vice versa during the copy procedure with a
fixed probability, denoted by $R$.
This probability is called the \emph{error rate}, or, alternatively, the
\emph{mutation rate}.

Molecules are also subject to decay with a particular rate $D_i$ for species
$i$. A molecule that decays is assumed to be completely absorbed by the
environment, i.e., it does not break into parts that are themselves being
described by the model.

The constant production of new molecules due to the ongoing
self-replication will drastically increase the concentration of
molecules, and will decrease the amount of resources available
for further self-replication. We are interested in the
description of a steady state, and therefore we have to introduce a constant
dilution which lets new resources enter the system and washes
away the excess production of those molecules. The total
concentration of molecules can thus be kept constant by proper
adjustment of the dilution flux.

Finally, we assume that the self-replicating molecules are placed in a
well-stirred reactor. As a consequence of this assumption, we can
neglect any spatial correlations in the model, and concentrate
solely on the molecules' abundances.

In summary, the model is based on the following postulates:
\begin{enumerate}
\item The molecules are represented by binary sequences of length
  $l$. They form and decompose steadily. The number of copies of
  sequence $i$ present at time~$t$ is denoted by $n_i(t)$.

\item Sequences enter the system solely as the result of a correct
  or erroneous copy of another sequence already present.

\item\label{item:quasispecies-ass3} The substrates necessary for the
  ongoing replication are always
  present in sufficient quantity. Excess molecules are washed
  away by a flux $\Phi (t)$.

\item The decay of sequences is a Poisson process.
\end{enumerate}

These four assumptions form the basis of Eigen's theory of molecular
evolution. A quantitative analysis of these assumptions can be done in
terms of differential equations in the molecules'
occupation numbers $n_i(t)$. In the following paragraphs, we
recall the quantitative analysis of the model that has been developed
by Eigen and coworkers.

We begin by writing down an expression for the change in the number of
copies of sequence $i$. The abundance $n_i(t)$ of sequence $i$
increases because a proportion of the molecules of type $i$ replicates
faithfully, while some of the other molecules of other types produce
offspring of type $i$ as the outcome of fruitless attempts to self-replicate.
Let the matrix $\mat Q = \big(Q_{ij}\big)$ express the
probability that molecule $j$ copies into molecule $i$. The associated
increase in the
abundance $n_i(t)$ then amounts to $\sum_j A_jQ_{ij}
n_j(t)$.

The decrease of a
molecule's abundance can have two reasons: its decay, expressed by
$-D_i n_i(t)$, and its removal due to the
flux term, expressed by $-n_i(t)\Phi(t)/N(t)$. Here, $N(t)$ is the total
number of molecules at time $t$, i.e., $N(t)=\sum_j n_j(t)$. Putting all
the different terms together, we end up with the net change $\dot n_i(t)$ in
the number of copies of sequence $i$,
\begin{equation}\label{eq:quasispecies1}
  \dot n_i(t)= \Big[ A_i Q_{ii} - D_i\Big] n_i(t)
     + \sum_{j\neq i} A_j Q_{ij} n_j(t) -
     \frac{n_i(t)}{N(t)} \Phi(t)\,.
\end{equation}
The quantity $Q_{ii}$ gives the probability that a molecule $i$
replicates faithfully. $Q_{ii}$ is sometimes referred to as the
\emph{copy fidelity}.

For the further development of the theory, it is
useful to introduce an average excess production $\bar E(t)$,
defined by
\begin{equation}\label{eq:def-excessprod}
  \bar E(t) := \sum_i n_i(t)\Big[A_i- D_i\Big]\Big/N(t)\,.
\end{equation}
The conservation law
\begin{equation}\label{eq-conserveQ}
  \sum_i Q_{ij} = 1\,,
\end{equation}
expressing the fact that every (possibly erroneous) copy of a molecule
represents another molecule in the chemistry, allows us to rewrite the average
excess production in terms of the total production rate and the dilution
flux, viz.
\begin{equation}\label{eq:excessprod-alt}
  \bar E(t)= \left[\sum_i \dot n_i(t)+\Phi(t)\right]\Big/ N(t) \,.
\end{equation}
It proves to be useful to define the
matrix $\mat W=\big(W_{ij}\big)$ as
\begin{equation}\label{eq:defWij}
  W_{ij} := A_j Q_{ij} - D_{i} \delta_{ij}\,.
\end{equation}
Now we can rewrite Eq.~(\ref{eq:quasispecies1}) as
\begin{equation}\label{eq:quasispecies2}
  \dot n_i(t)= \Big[ W_{ii} - \bar E(t)\Big] n_i(t)
      + \sum_{j\neq i} W_{ij} n_j(t) + n_i(t)
              \frac{\dot N(t)}{N(t)}\,.
\end{equation}
Let us introduce concentration variables $x_i(t)$, defined by
\begin{equation}
  x_i(t) := \frac{n_i(t)}{N(t)}\,.
\end{equation}
The quantity $x_i(t)$ measures the relative concentration of molecule $i$ in
the population. The change in $x_i(t)$ is given by
\begin{equation}
  \dot x_i(t) = \frac{\dot n_i(t)}{N(t)} - n_i(t) \frac{\dot N(t)}{N(t)^2}\,.
\end{equation}
Hence, if we subtract the rightmost term of Eq.~(\ref{eq:quasispecies2}) on both
sides of Eq.~(\ref{eq:quasispecies2}), and divide by $N(t)$, we arrive at
\begin{equation}\label{eq:quasispecies-conc-novec}
 \dot x_i(t)=\Big[W_{ii}-\bar E(t)\Big] x_i(t) + \sum_{j\neq i}
 W_{ij}(t) x_j(t)\,.
\end{equation}
The total number of molecules grows with
\begin{equation}
  \dot N(t) = \bar E(t)N(t)-\Phi(t)\,.
\end{equation}
Typically, one assumes that the flux $\Phi(t)$ is adjusted such that $\dot
N(t)$ vanishes, as expressed by Assumption~\ref{item:quasispecies-ass3} on
page~\pageref{item:quasispecies-ass3}. However, this assumption is not really
necessary for the further development of the theory. Since from this point
onwards, we will only be concerned with the relative concentrations $x_i(t)$,
we will disregard the flux altogether, and work with
Eq.~(\ref{eq:quasispecies-conc-novec}) exclusively.

At this point, it is useful to introduce vector notation, by lumping
the concentrations $x_1(t), x_2(t), \dots$ together into a single vector
$\vec x(t)=\big(x_1(t), x_2(t), \dots\big)$.
Equation~(\ref{eq:quasispecies-conc-novec}) then becomes
\begin{equation}\label{eq:quasispecies-conc}
 \dot{\vec x}(t)=\Big[\mat W-\bar E(t)\Idmat\Big] \vec x(t),
\end{equation}
where $\Idmat$ is the identity matrix. The matrix $\mat W$ can be
decomposed into
\begin{equation}
\mat W = \mat Q \mat A - \mat D\,,
\end{equation}
if we write the replication and the decay coefficients in matrix
form as well. Both $\mat A$ and $\mat D$ are diagonal matrices of the form
\begin{subequations}\label{eq:matrixAD}
\begin{align}
  \mat A &= \diag \left(A_1, A_2, \dots \right)\,,\\
  \mat D &= \diag \left(D_1, D_2, \dots \right)\,.
\end{align}
\end{subequations}
The matrix $\mat Q$ is the mutation matrix introduced above. Note that the
average excess production can also be transformed into vector notation. It
takes on the form
\begin{equation}
  \overline E(t) = \vec e^\transp\cdot [\mat A(t)\vec x(t) - \mat D(t)\vec
  x(t)]\,,
\end{equation}
where $\vec e^\transp$ is a vector of 1s, i.e.,
$\vec e^\transp = (1,\dots, 1)$. The scalar product between $\vec e^t$ and a
concentration vector [say $\vec y(t)$]
gives exactly the sum over all components of that vector.

Equation (\ref{eq:quasispecies-conc}) is nonlinear, since the term $\bar
E(t) \vec x(t)$ is quadratic in $\vec x(t)$. Nevertheless,  a
straightforward solution of Eq~(\ref{eq:quasispecies-conc}) is possible,
because a transformation exists which removes the nonlinearity. The
strength of this transformation lies in the easy reconstruction of the
concentration variables $x_i(t)$ from the transformed
system. Following Thompson and McBride~\cite{ThompsonMcBride74}, or
Jones \etal~\cite{JonesEnnsRangnekar76}, we introduce
\begin{equation}\label{eq:xtoy-nonl-trans}
  \vec y(t):=\exp\left( \int_0^t \bar E(\tau)\, d\tau\right) \vec x(t)\,.
\end{equation}
The new variables satisfy the linear equation
\begin{equation}\label{eq:lin-quasispecies}
  \dot{\vec y}(t)=\mat W \vec y(t)\,,
\end{equation}
which can be shown by insertion of Eq.~(\ref{eq:xtoy-nonl-trans})
into Eq.~(\ref{eq:lin-quasispecies}). Moreover, since $\vec y(t)$
differs from $\vec x(t)$ only by a scalar factor, we can restore
the original variables via
\begin{equation}\label{eq:x-of-y}
  \vec x(t)=\frac{\vec y(t)}{\vec e^\transp\cdot\vec y(t)}\,.
\end{equation}
Note that if all decay constants are equal, i.e., $\mat
D=\diag (D,\dots,D)$ with a single scalar value $D$, then the
transformation
\begin{equation}\label{eq:extended-lin-trans-static}
  \vec y(t)=\exp\left(\int_0^t[ \overline E(\tau)+ D]\,d\tau\right)\vec x(t)
\end{equation}
leads to the even simpler equation
\begin{equation}\label{eq:simplest-linearized-eq-static}
  \dot{\vec y}(t) = \mat Q\mat A\vec y(t)\,.
\end{equation}
The concentration vector $\vec x(t)$ can again be obtained from
Eq.~(\ref{eq:x-of-y}).

The mutation matrix $\mat Q$ has so far remained undefined. The choice
of $\mat Q$ specifies the relationship between the different molecules
in the chemistry. As was stated above, we conceive the
molecules as bitstrings of length $l$, and we assume that copy errors
are equally likely on all positions on the string. In that case, the
mutation matrix can be written down straightforwardly. Suppose that
two bitstrings
differ in $d$ positions. A mutation from one to the other occurs only if
exactly these $d$ positions are copied erroneously, while all others
are copied faithfully. Such a copy error appears with probability
$(1-R)^{l-d}R^d$. Hence, we can write the mutation matrix $\mat Q$ as
\begin{equation}\label{eq:def-Q}
  Q_{ij}=(1-R)^l\left(\frac{R}{1-R}\right)^{d(i,j)},
\end{equation}
where $d(i,j)$ is the Hamming distance between two sequences. The
Hamming distance is the number of bits in which two sequences differ.

The matrix $\mat Q$ is typically very large, since its number of rows
and columns grows as $2^l$ with increasing sequence length $l$. This
makes Eq.~(\ref{eq:lin-quasispecies}) almost intractable, numerically
as well as analytically, for all but the shortest
sequences. An
alternative matrix $\mat Q'$ is often used, in which certain
sequences are grouped together, such that the number of concentration
variables can be reduced to $l+1$. The main idea of this grouping,
developed by Swetina and Schuster~\cite{SwetinaSchuster82},
is to define a single sequence (which should have the highest
replication coefficient of all sequences) as master sequence, and to
group all other sequences into error classes, according
to their
Hamming distance from the chosen master sequence. All the sequences
with the same Hamming distance from the master form a single error
class. This procedure has
the advantage of greatly reducing the number of variables, but it also
restricts considerably the number of fitness landscapes that can
be studied. Since all
sequences in an error class have to share the same replication
coefficient, a fitness landscape, for example, in which two peaks have a small
or moderately large Hamming distance cannot be defined.

The error class matrix $\mat Q'$ has been given by Nowak and
Schuster~\cite{NowakSchuster89} in a relatively simple form:
\begin{equation}\label{eq:Qswetschust}
  Q'_{ij}=\sum_{k=\max\{i+j-l,0\}}^{\min\{i,j\}} \binom{j}{k}\binom{l-j}{i-k}(1-R)^l\left(\frac{R}{1-R}\right)^{i+j-2k}\,.
\end{equation}
Note that in \cite{NowakSchuster89}, the indices $i$ and $j$ are
interchanged inadvertently. The error class matrix can be
derived from Eq.~(\ref{eq:def-Q}) by deletion of all but one column
of every group of columns whose index $j$ yields the same $d(0,j)$,
and by the subsequent summation of all rows whose index $i$ yields the same
$d(i,0)$.

The linearized evolution
equation~(\ref{eq:lin-quasispecies}) can be solved analytically. The
transition from an initial state $\vec y(0)$ to the state at time $t$ is given
by~\cite{Erugin66}
\begin{equation}\label{eq:formal-solution}
 \vec y(t) =\exp(\mat W t)\vec y(0)\,.
\end{equation}
From that expression, we can read off that the (unnormalized) variables $\vec
y(t)$ will grow exponentially over time. Mathematically, this growth can be
accompanied by either exponentially damped or exponentially amplified
oscillations. For all cases of biological interest, however, there
can be at most exponentially damped oscillations. First of all, we will almost
always deal with a symmetric matrix $\mat Q$. A symmetric mutation
matrix implies that the probability with which a sequence $i$ mutates into a
sequence $j$ is the same as the probability for the inverse
process. Rumschitzki has noted that in this case, the spectrum of $\mat W$ is
real~\cite{Rumschitzki87}. Namely, we can transform the non-symmetric matrix
$\mat W$ into a symmetric one by means of a similarity transformation,
\begin{equation}
  \mat W = \mat Q\mat A - \mat D \quad \rightarrow \quad
  \mat A^{1/2}\mat W\mat A^{-1/2} =  \mat A^{1/2}\mat Q\mat A^{1/2} -
  \mat D \,.
\end{equation}
The spectrum of the transformed matrix is real because of the matrix's
symmetry, and hence the spectrum of the untransformed matrix must be real as
well. In that case, oscillations will be absent in
Eq.~(\ref{eq:formal-solution}). For non-symmetric $\mat Q$, we can apply the
Frobenius-Perron theorem~\cite{Perron07}
if the decay rates satisfy
\begin{equation}\label{eq:Frob-Perron-cond}
  (\mat D)_{ii} < (\mat Q\mat A)_{ii}  \qquad\mbox{for all $i$.}
\end{equation}
The Frobenius-Perron theorem is applicable to a matrix with all positive
entries, and it guarantees a real largest
eigenvalue for that matrix. Consequently, we have at most exponentially damped
oscillations as long as (\ref{eq:Frob-Perron-cond}) is obeyed.
In addition to that, the Frobenius-Perron theorem states that the
eigenvector corresponding to this largest eigenvalue has only strictly
positive entries, and hence, that this eigenvector can be interpreted
as a vector of chemical concentrations if normalized
appropriately.

Let us write down a more explicit solution to
\eqref{eq:lin-quasispecies}, under the assumption that the spectrum of
$\mat W$ is known (exact spectra of evolution matrices have been derived
in~\cite{Rumschitzki87,DressRumschitzki88,Galluccio97}). Let $\lambda_k$ be
the eigenvalues of
$\mat W$, and let $\vec\phi_k$ be
the associated eigenvectors.  Without loss of generality, we order the
eigenvalues such that
$\lambda_0\geq\lambda_1\geq\lambda_2\geq\dots$\,. The solution
to Eq.~(\ref{eq:lin-quasispecies}) can then be expressed as
\begin{equation}\label{eq:sol-y-Aconst}
  \vec y(t) = \sum_{k} \alpha_k \vec \phi_k\, e^{\lambda_k t}\,,
\end{equation}
where the coefficients $\alpha_k$ have to be chosen such that the
initial condition $\vec y(t=0)=:\vec y_0$ is satisfied. In other words,
the $\alpha_k$ give the expansion of the initial condition $\vec y_0$
in terms of the eigenvectors $\vec \phi_k$,
\begin{equation}
  \vec y_0 = \sum_k \alpha_k \vec \phi_k\,.
\end{equation}
The principal eigenvector $\vec \phi_0$ of $\mat W$ has been called the
quasispecies by Eigen. The reason for this name will become clear shortly.

In terms of the concentration variables $\vec x(t)$, the solution
Eq.~(\ref{eq:sol-y-Aconst}) becomes
\begin{equation}\label{eq:sol-x-Aconst}
  \vec x(t)=\frac{\vec y(t)}{\vec e^\transp\cdot \vec y(t)}\,,
\end{equation}
where $\vec e$ is again a vector with all entries
equal to 1.

As we already mentioned, the Perron-Frobenius
theorem assures that the largest eigenvalue $\lambda_0$ is
non-degenerate, and that all components of the corresponding
eigenvector $\vec \phi_0$ are positive.
We factor out the
exponential of the largest eigenvalue $e^{\lambda_0 t}$ in the
  numerator and denominator of Eq.~(\ref{eq:sol-x-Aconst}) and obtain
\begin{equation}\label{eq:expsol-x-Aconst}
  \vec x(t)=\frac{\alpha_0\vec\phi_0 + \sum_{k>0}\alpha_k\vec\phi_k\,
                      e^{(\lambda_k-\lambda_0)t}}
   {\sum_{k}\alpha_k\,\vec e\cdot\vec\phi_k\, e^{(\lambda_k-\lambda_0)t}}\,.
\end{equation}
As long as $\alpha_0\neq 0$, all contributions apart from the
one corresponding to the largest eigenvalue disappear in the limit
$t\rightarrow\infty$. Hence, the
system evolves towards a steady state, characterized by the dominant
eigenvector of the matrix $\mat W$.

The case $\alpha_0=0$ is somewhat artificial, because it implies that the
system has been started in an exact superposition of eigenstates
excluding the dominant quasispecies. In that case, the system cannot
evolve towards $\vec \phi_0$. Instead, it converges towards the
eigenvector corresponding to the next-largest eigenvalue. In a real
chemical system with more or less random initial conditions and under
the presence of noise, the case $\alpha_0=0$ is of no relevance.

The appearance of a steady state distribution of concentration
variables has important implications for the understanding of
Darwinian evolution. In general, the eigenvectors $\vec\phi_k$ are a
mixture of several molecular species~$i$.  As a consequence,
a number of species is present simultaneously in the asymptotic
distribution. This means that selection combined with random
mutation does not remove all but one molecular species, even if there
exists one (the master sequence) whose replication coefficient exceeds
all others. Instead, the interplay of selection and mutation creates a
\emph{cloud} of mutant species around the master sequence. It is this cloud
that is termed the quasispecies. The mutant distribution forms a unit
competing with other similar units, represented by the other
eigenvectors. Selection acts on these mixtures of sequences, and ultimately
singles out the dominant one, i.e., the one corresponding to the largest
eigenvalue.

\begin{figure}[tb]
\centerline{
 \epsfig{file={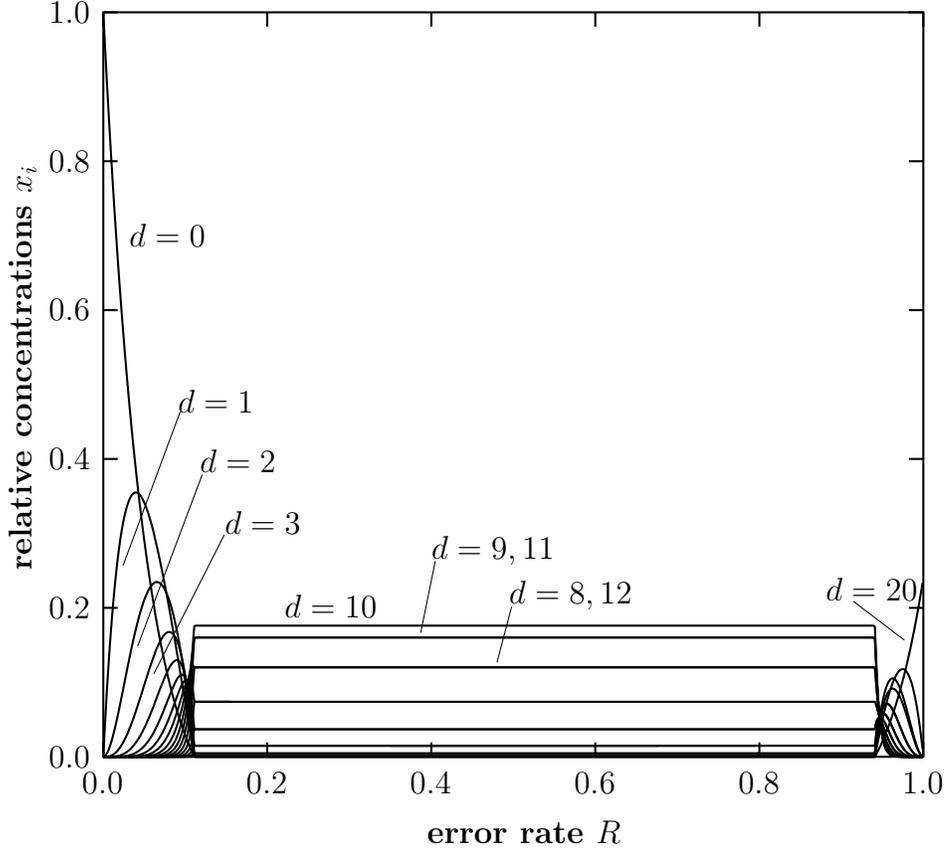}, width=.9\columnwidth}
}
\caption{The relative concentrations of the master sequence and the
 error classes for sequences of length $l=20$ versus the error rate~$R$ in a
 sharply peaked landscape with replication coefficients $A_0=10, A_1, \dots,
 A_{20}=1$. The concentrations of all
 sequences with the same distance~$d$ to the master are summed up and
 displayed as a single line. The decay constants have been set to
 $D_i=D$, with arbitrary~$D$. In static fitness landscapes, the decay
 constants drop out of the asymptotic quasispecies if they are all equal.
\label{fig:qs:fixedSwetinaSchuster}
}
\end{figure}

Interestingly, the master sequence does not necessarily
occupy a large fraction of the quasispecies. In fact, as the mutation rate
increases, it is common that the master dwindles while the one-mutant
and the two-mutant sequences form the prevailing part of the
quasispecies. Figure~\ref{fig:qs:fixedSwetinaSchuster} shows the asymptotic
sequence distribution versus the error rate $R$
for a sharply peaked landscape in which all but one of the
replication coefficients are identical, while the remaining one exceeds
them significantly. The mutants are grouped into error classes. If
copy errors are not present, i.e., for $R=0$, the master sequence is
the only species present, and all others have a vanishing
concentration. As soon as $R$ takes on values slightly above~0, the
concentration of the master sequence starts to decrease. At the same
time, the concentrations of the mutants begin to grow, first that
of the one-mutant sequence, than that of the two-mutant sequence,
and so on. At $R=0.11$ we observe a sharp transition, beyond which the
concentrations take on constant values. The mutation rate at which this
transition occurs is called the \emph{error threshold}. Beyond the error
threshold, the outcome of a replication event can be considered a random
sequence, and
therefore, all molecular species take on the same concentration in this
regime. The fact that all concentrations are equal beyond the error threshold
is somewhat blurred in Fig.~\ref{fig:qs:fixedSwetinaSchuster} because of
the use of error classes. In fact, the heights of the different
curves merely reflect the sizes of the corresponding error classes.
For $R=\frac{1}{2}$, the sequences replicate stochastically in any
landscape, since faithfully copied and erroneously copied symbols are then
equally likely. The outcome of a single copy process is therefore a
random sequence in that case. When~$R$ comes close to the value~1,
almost every symbol is copied incorrectly. This implies that the copy is,
apart from mutations, the inverse of the original
sequence (note that this is a peculiarity of binary strings). As a
consequence, order can again be seen for large~$R$. In
Fig.~\ref{fig:qs:fixedSwetinaSchuster}, the inverse error
threshold occurs at $R=0.94$. Beyond that point, the inverse master
sequence dominates the quasispecies.

If an evolving system displays an error threshold, the amount of information
that can be acquired by the system is severely limited. In the case of the
sharply peaked landscape, the critical mutation rate at which the error
threshold occurs decreases as $1-(1/\sigma)^{1/l}$ with increasing string
length~\cite{MaynardSmith83}. The quantity $\sigma$ in that expression stands
for the relative advantage of the master sequence over the other
sequences. For large $l$, the critical mutation rate is therefore very close
to 0. This implies that for a given mutation rate, the
molecules can grow to a certain length, while beyond that length, all
information is lost. Incidentally, it follows that the spontaneous formation
of complex
self-replicating molecules seems to be impossible. Eigen and Schuster attempted
to solve this problem with the concept of the
hypercycle~\cite{EigenSchuster79}. However, it must be said that not all
landscapes do present an error threshold. In particular, multiplicative
landscapes~\cite{Higgs94,WoodcockHiggs96,WagnerBaakeGerisch97} show a smooth
crossover from the completely ordered situation at $R=0$, where the master
sequence is the only sequence present, to $R=\frac{1}{2}$, where all molecules
take on the same concentration. Moreover, the decay of the master sequence's
concentration depends only on the mutation rate and the selective advantage,
but not on the length of the strings. Hence, in a multiplicative model,
sequences can grow in length without bound, and self-replicating molecules
might form spontaneously.

The error threshold can be viewed as the critical point of a phase
transition. This fact is well
established through a rich body of work, in which the quasispecies model and
related models have been mapped onto spin
lattices~\cite{Leuthaeusser85,Leuthaeusser87,Tarazona92}, spin
chains~\cite{Baakeetal97,WagnerBaakeGerisch97}, and more recently also onto
the statistical mechanics of directed polymers in a random
medium~\cite{Galluccio97}. The order parameter that is generally being used to
describe that phase transition is the average overlap of the sequences in the
population with the master sequence, given by the expression
$l-2d(0,i)$, where $d(0,i)$ is the Hamming distance between a sequence of type
$i$ and the master sequence. The overlap takes on the value $l$ if we compare
the master sequence with itself, the value $-l$ if we compare
it with its inverse, and intermediate values for all other sequences. The
order parameter reads
\begin{equation}\label{eq:def-order-parameter}
  m_s = \frac{1}{l} \sum_i x_i[l-2d(0,i)]\,.
\end{equation}
We use the symbol $m_s$ because the order parameter represents the surface
magnetization when we map the quasispecies model onto an Ising
lattice~\cite{Leuthaeusser87,Tarazona92}.

\subsection{Time-dependent replication rates}
\label{sec:time-dependent-rep-rates}

Having developed a good understanding of the concepts of molecular
evolution theory in static environments, let us move on to the
non-static case. The basic quasispecies equation~(\ref{eq:quasispecies-conc})
changes only in so far as the matrix $\mat W$
now becomes time dependent,
\begin{equation}\label{eq:basic-quasispecies}
  \dot {\vec x}(t)=[\mat W(t)-\overline E(t)\Idmat] \vec x(t)\,.
\end{equation}
In the most general case, the time dependency may stem from the replication
rates, from the decay rates, or even from the mutation matrix:
\begin{equation}
  \mat W(t)=\mat Q(t)\mat A(t)-\mat D(t)\,.
\end{equation}
The average excess production is then
\begin{equation}
  \overline E(t) = \vec e^\transp\cdot [\mat A(t)\vec x(t) - \mat D(t)\vec
  x(t)]\,.
\end{equation}
As in the static case, we can transform away the nonlinearity in
Eq.~(\ref{eq:basic-quasispecies}) with the introduction of unnormalized
variables
\begin{equation}\label{eq:jones-trans}
  \vec y(t)=\exp\left(\int_0^t \overline E(\tau)\,d\tau\right)\vec x(t)\,.
\end{equation}
The resulting equation is again a linear differential equation, however, this
time with non-constant coefficients:
\begin{equation}\label{eq:basic-linearized-eq}
  \dot{\vec y}(t) = \mat W(t)\vec y(t)\,.
\end{equation}
As in the static case, if all decay constants are equal, i.e.,
$\mat D(t)=\diag (D(t),\dots,D(t))$ with a single scalar function
$D(t)$, then an extended transformation similar to
Eq.~(\ref{eq:extended-lin-trans-static}) removes the decay constants from
Eq.~(\ref{eq:basic-linearized-eq}).

In the previous subsection, we were able to immediately write down a solution
for the linearized quasispecies equation. In the case of a dynamic fitness
landscape, however, no such a simple closed form solution exists. Instead, we
have to be satisfied with partial solutions, approximations, and
limiting cases. In particular, we cannot generally define a quasispecies as
the steady state the system approaches for $t\rightarrow\infty$. Therefore, a
central question in relation to dynamic landscapes is the appropriate
definition of a quasispecies concept.

For now, we start collecting information about
Eq.~(\ref{eq:basic-linearized-eq}). First of all, we note that we can
map the quasispecies model onto a linear system with a symmetric
matrix $\tilde{\mat W}(t)$ if $\mat Q(t)$ is symmetric for all $t$.
This can be seen by introducing
\begin{equation}
  \vec z(t) = \mat A^{1/2}(t) \vec y(t)\,.
\end{equation}
Differentiation yields
\begin{gather}\label{eq:symmetric-quasisp-eq}
  \dot{\vec z}(t) = \tilde{\mat W}(t) \vec z(t)
\end{gather}
with
\begin{equation}
\tilde{\mat W}(t) = \mat A^{1/2}(t)\mat Q(t)\mat A^{1/2}(t)
-\mat D(t)+\left[ \frac{d}{dt}\mat A^{1/2}(t)\right]\mat A^{-1/2}(t)\,.
\end{equation}
Unfortunately, we cannot write down a solution for
Eq.~(\ref{eq:symmetric-quasisp-eq}) from the knowledge of the
eigensystem of $\tilde{\mat W}(t)$ if $\tilde{\mat W}(t)$ has an
arbitrary time dependency. However, the above mapping shows that for symmetric
$\mat Q(t)$, we may safely assume that the matrix $\mat W(t)$ has only real
eigenvalues.

There are two limiting cases that we can discuss without specifying a
landscape, namely very fast changes in $\mat W(t)$
on the one hand, and very slow changes in $\mat W(t)$
on the other hand. We begin with the case of very slow changes. For
the rest of this work, we will assume that $\mat W(t)$ has a real
spectrum for all $t$. To be on the safe side, we also assume that
(\ref{eq:Frob-Perron-cond}) is satisfied for all $t$. In that way, the
Perron eigenvector of $\mat W(t)$ can always be interpreted as a
vector of chemical concentrations.

For every time $t_0$, we can define a relaxation time
\begin{equation}
  \tau_{\rm R}(t_0) = \frac{1}{\lambda_0(t_0)-\lambda_1(t_0)}\,,
\end{equation}
where $\lambda_0(t)_0$ and $\lambda_1(t_0)$ are the largest and the
second largest eigenvalue of $\mat W(t_0)$, respectively. The time
$\tau_{\rm R}(t_0)$ gives an estimate of how long a linear system with matrix
$\mat W(t_0)$ needs to settle into equilibrium. Therefore, if the
changes in $\mat W(t)$ happen on a timescale much longer than
$\tau_{\rm R}(t)$, the system is virtually in equilibrium at any given point
in time. Hence, for large enough $t$, the quasispecies will be given
by the Perron eigenvector of $\mat W(t)$. Strictly speaking, this is
only true if there is always some overlap between the largest
eigenvector of $\mat W(t)$ and the one of
$\mat W(t+dt)$, but in all but some very pathological cases we can
assume this to be the case.

The situation of fast changes in $\mat W(t)$ is somewhat more
difficult, because, as we are going to see later on, we have to define a
suitable average over $\mat W(t)$ in order to make a general
statement. Therefore, we postpone this case for a moment. A
detailed discussion of fast changes will be given for the particular
case of periodic fitness landscapes in the next section, and later on,
we will discuss fast changing landscapes in general.

\section{Periodic fitness landscapes}
\label{sec:periodic-fitness-landscapes}
\subsection{Differential equation formalism}
\label{sec:diff-equ-formalism}

In this section, we study periodic time dependencies in
$\mat W(t)$, for which we can prove several general
statements.

If the changes in $\mat W(t)$ are periodic, i.e., if there exists a
$T$ such that
\begin{equation}
  \mat W(t+T)= \mat W(t) \quad\mbox{for all $t$,}
\end{equation}
then Eq.~(\ref{eq:basic-linearized-eq}) turns into a system of linear
differential equations with periodic coefficients. Several
theorems are known for such
systems~\cite{YakubovichStarzhinskii75}. Most
notably, if $\mat Y(t, t_0)$ is the fundamental matrix, such that
every solution to Eq.~(\ref{eq:basic-linearized-eq}) can be written in
the form
\begin{equation}\label{eq:lin-solution}
  \vec y(t)=\mat Y(t,t_0)\vec y(t_0)\,,
\end{equation}
then we can define the \emph{monodromy} matrix $\mat X(t_0)$,
\begin{equation}\label{eq:def-monodromy}
  \mat X(t_0) = \mat Y(t_0+T,t_0)\,,
\end{equation}
which simplifies Eq.~(\ref{eq:lin-solution}) to
\begin{align}\label{eq:gen-periodic-sol}
  \vec y(t)&= \mat Y(t_0+\phi, t_0)\mat X^m(t_0)\vec y(t_0)\notag\\
  &=\mat X^m(t_0+\phi) \mat Y(t_0+\phi, t_0)\vec y(t_0)\,,
\end{align}
for the decomposition $t=mT+\phi+t_0$ with the phase $\phi<T$. In
particular, we have
\begin{equation}\label{eq:basic-mon-solution}
  \vec y(\phi+mT)= \mat X^m(\phi)\vec y(\phi)\,,
\end{equation}
so that for every phase $\phi$, we
have a well defined asymptotic solution, given by the eigenvector associated
with the largest eigenvalue of $\mat X(\phi)$. Hence, periodic
fitness landscapes allow the definition of a quasispecies,
much in the same way as static fitness landscapes do. However, here the
quasispecies is time-dependent. In other words, a system that evolves in a
periodic fitness landscape runs for $t\rightarrow\infty$ into a limit cycle
with period length $T$.

\subsubsection{Neumann series for $\mat X$}

With the knowledge of the monodromy matrix $\mat X$, the calculation of the
system's limit cycle becomes a simple eigenvalue problem. The monodromy
matrix, however, can in general not be given in a closed from. Consequently,
we have to rely on expansions in various parameters. In this section, we
derive a formal expansion in $T$ for the monodromy matrix. This
formal expansion is similar in spirit to the Neumann series which
gives a formal solution to an integral equation, and it is based on
the Picard-Lindel\"of iteration for differential equations.
As the first step, we have to rewrite
Eq.~(\ref{eq:basic-linearized-eq}) in the form of an integral
equation, i.e.,
\begin{equation}\label{eq:integral-equation}
  \vec y(t_0+\tau) =\vec y(t_0) + \int_0^\tau \mat W(t_0+\tau_1)\vec y(t_0+\tau_1)d\tau_1\,.
\end{equation}
Our goal is to solve this equation for $\vec y(t_0+\tau)$ by iteration. Our
initial solution is
\begin{equation}
  \vec y_0(t_0+\tau) = \vec y(t_0)\,,
\end{equation}
which we insert into Eq.~(\ref{eq:integral-equation}). As a result, we
obtain the 1st order approximation
\begin{equation}
  \vec y_1(t_0+\tau) = \vec y(t_0) + \int_0^\tau \mat W(t_0+\tau_1)\vec
  y(t_0)d\tau_1\,.
\end{equation}
Further iteration yields
\begin{multline}
  \vec y_2(t_0+\tau) = \vec y(t_0) + \int_0^\tau \mat W(t_0+\tau_1)\vec
  y(t_0)d\tau_1 \\
   + \int_0^\tau  \mat W(t_0+\tau_1) \int_0^{\tau_1} \mat
  W(t_0+\tau_2)\vec  y(t_0)d\tau_1d\tau_2\,,
\end{multline}
and so on. Now we define
\begin{align}
  \overline {\mat W}_0(t_0, \tau)&=1\,,\\
  \overline {\mat W}_1(t_0, \tau)&=\frac{1}{\tau}\int_0^\tau \mat W(t_0+\tau_1)d\tau_1\,,
\end{align}
and, in general
\begin{equation}\label{eq:w-av-arbitrary-order}
  \overline {\mat W}_k(t_0,\tau) =\frac{1}{\tau^k}\int_0^\tau \mat W(t_0+\tau_1)
      \int_0^{\tau_1} \mat W(t_0+\tau_2)\cdots\int_0^{\tau_k-1} \mat W(t_0+\tau_k)
         d\tau_1 d\tau_2 \cdots d\tau_k\,,
\end{equation}
and obtain the formal solution
\begin{equation}\label{eq:periodic-formal-sol}
  \vec y(t_0+\tau) =  \sum_{k=0}^\infty \tau^k \overline{\mat W}_k(t_0,\tau)\vec y (t_0)\,.
\end{equation}
For suitably small $\tau$, the infinite sum on the right-hand side is
guaranteed to converge.
When we compare this equation for $\tau=T$ to the definition of the monodromy
matrix Eq.~(\ref{eq:def-monodromy}), we find that [introducing
$\overline{\mat W}_k(t_0):=\overline{\mat W}_k(t_0,T)$]
\begin{equation}\label{eq:X-Neumann-exp}
  \mat X(t_0) = \sum_{k=0}^\infty T^k \overline{\mat W}_k(t_0)\,.
\end{equation}
In particular, since $\overline{\mat W}_1(t_0)$ is identical to the time-average
over $\mat W(t)$, regardless of $t_0$, we have the high-frequency
expansion
\begin{equation}\label{eq:high-freq-expans}
  \mat X(t_0) = \Idmat + T\overline{\mat W} + {\mathcal{O}}(T^2)\,,
\end{equation}
with
\begin{equation}
  \overline{\mat W}=\frac{1}{T}\int_0^T \mat W(t)\,dt\,.
\end{equation}
Equation (\ref{eq:high-freq-expans}) reveals that for very high
frequency oscillations, the system behaves as if it was subject to
a static landscape. That static landscape is given by the dynamic
landscape's average over one oscillation period.

The radius of convergence of the expansion
Eq.~(\ref{eq:X-Neumann-exp}) can be estimated as follows. Since all
entries of $\mat W(t)$ are positive, the tensor
\begin{align}
  \overline W_{i\nu_1}\overline W_{\nu_1\nu_2}\cdots\overline W_{\nu_{k-1}j}(t_0)&:=
  \frac{1}{T^k}\int_0^T W_{i\nu_1}(t_0+\tau_1) \int_0^{\tau_1}
  W_{\nu1\nu_2}(t_0+\tau_2)
  \cdots\notag\\
&\qquad \qquad\int_0^{\tau_{k-1}} W_{\nu_{k-1}j}(t_0+\tau_k)
         d\tau_1d\tau_2\cdots d\tau_k
\end{align}
can be bound by
\begin{align}
  \overline W_{i\nu_1}\overline W_{\nu_1\nu_2}\cdots\overline W_{\nu_{k-1}j}(t_0)&\leq
  \frac{1}{T^k}\int_0^T W_{i\nu_1}(t_0+\tau)d\tau \int_0^{T}
  W_{\nu1\nu_2}(t_0+\tau)d\tau
  \cdots\notag\\
&\qquad\qquad \int_0^{T} W_{\nu_{k-1}j}(t_0+\tau) d\tau\,,
\end{align}
from which it follows that
\begin{equation}\label{eq:estimate-W-entries}
  \left( \overline{\mat W}_k(t_0)\right)_{ij} \leq \left( \overline{\mat W^k}\right)_{ij}\,.
\end{equation}
The matrix norm induced by the sum norm
\begin{equation}
  \norm{(y_1, y_2, \dots, y_n)}_1 = \sum_i |y_i|
\end{equation}
is the column-sum norm
\begin{equation}
  \norm{\overline{\mat W}}_1 = \max_{j}\left\{\sum_i |\overline W_{ij}|\right\}\,.
\end{equation}
With that norm, and using Eq.~(\ref{eq:estimate-W-entries}), we can estimate
\begin{equation}
  \norm{\overline{\mat W}_k(t_0)}_1\leq \norm{\overline {\mat W}^k}_1 \leq
         \norm{\overline {\mat W}}_1^k\,.
\end{equation}
Hence, the expansion Eq.~(\ref{eq:X-Neumann-exp}) converges certainly
for those $T$ that satisfy
\begin{equation}\label{eq:convergence-crit}
  T \norm{\overline{\mat W}}_1 < 1\,.
\end{equation}
Since all entries in $\mat W$ are positive, we have further
\begin{align}
  \norm{\overline{\mat W}}_1 &= \max_{j} \left\{\sum_i |\overline A_j
    Q_{ij}-\overline D_j\delta_{ij}|\right\} \notag\\
  &= \max_{j} \left\{\overline A_j-\overline D_j\right\}\,,
\end{align}
where the bar in $\overline A_j$ and $\overline D_j$ indicates that these
quantities are averaged over one oscillation period. The second
equality holds because of (\ref{eq:Frob-Perron-cond}) and because of
$\sum_i \mat Q_{ij} = 1$.
Without loss of generality, we assume that the maximum is given by
$\overline A_0-\overline D_0$. Then, Eq.~(\ref{eq:convergence-crit})
is satisfied for
\begin{equation}\label{eq:radius-of-convergence}
  T < \frac{1}{\overline A_0-\overline D_0}\,.
\end{equation}
It is interesting to compare this expression to the relaxation time
of the time-averaged fitness landscape, $\overline\tau_{\rm R}$. To $0$th order,
the principal eigenvalue of $\overline{\mat W}$ is given by $\overline
W_{00}$. The second largest eigenvalue is to the same order given by
the second largest diagonal element of $\overline{\mat W}$, which we
assume to be $\overline W_{11}$ without loss of generality. Hence, the
relaxation time is approximately given by
\begin{equation}
  \overline\tau_{\rm R} = \frac{1}{\overline W_{00}-\overline W_{11}} >\frac{1}{\overline
  W_{00}}\geq \frac{1}{\overline A_0-\overline D_0} \,,
\end{equation}
which is in general larger than the radius of convergence of
Eq.~(\ref{eq:X-Neumann-exp}). In particular, if the largest and the
second largest eigenvalue of $\overline{\mat W}$ lie close together, the
relaxation time may be much larger than the largest oscillation period
for which the expansion is feasible. This restricts the usefulness of
Eq.~(\ref{eq:X-Neumann-exp}) to substantially high frequency
oscillations in the landscape. The interesting regime in which the
changes in the landscape happen on a time scale comparable to the
relaxation time of the system can unfortunately not be studied from
Eq.~(\ref{eq:X-Neumann-exp}).

\subsubsection{Exact solutions for $R=0$ and $R=0.5$}

The two extreme cases $R=0$ (no replication errors) and $R=0.5$
(random offspring sequences) allow for an exact analytic
treatment. The second case is identical to the case of static
landscapes, and therefore we will mention it only briefly. At the
point of stochastic replication $R=0.5$, the population dynamics
becomes independent of the details of the landscape. As a consequence,
temporal changes in the
landscape must become less important  as $R$ approaches
$R=0.5$. However, this is not very surprising, since in most cases, an
error rate close to 0.5 implies that the population has already passed
the error threshold, which in turn implies that it does not feel the
changes in the landscape any more.

The case of $R=0$, on the other hand, is more complex than the
corresponding case in a static landscape. Since the matrix $\mat Q$
becomes the identity matrix for $R=0$, Eq.~(\ref{eq:basic-linearized-eq})
reduces to
\begin{equation}\label{eq:basic-R0-eq}
  \dot{\vec y}(t) = [\mat A(t)-\mat D(t)]\vec y(t)\,.
\end{equation}
The matrices $\mat A(t)$ and $\mat D(t)$ are diagonal by definition,
and hence, a solution to Eq.~(\ref{eq:basic-R0-eq}) is given by
\begin{equation}\label{eq:sol-basic-R0-eq}
  \vec y(t) = \exp\left(\int_{t_0}^t [\mat A(t')-\mat D(t')]dt'\right)
               \vec y(t_0)\,.
\end{equation}
When we compare this expression to Eqs.~(\ref{eq:lin-solution})
and~(\ref{eq:def-monodromy}), we find
\begin{equation}
  \mat Y(t,t_0) = \exp\left(\int_{t_0}^t [\mat A(t')-\mat D(t')]dt'\right)\,,
\end{equation}
and, in particular,
\begin{equation}
  \mat X(\phi) = \exp\left(\int^{\phi+T}_\phi [\mat A(t')-\mat D(t')]dt'\right)\,.
\end{equation}
The integral in the second expression is taken over a complete
oscillation period, and hence,
it is independent of $\phi$. Thus, we find for arbitrary $\phi$
\begin{equation}
  \mat X(\phi) = \exp(\overline{\mat W})\quad\mbox{for $R=0$.}
\end{equation}
With a vanishing error rate, the monodromy matrix becomes the
exponential of the time-average over $\mat W(t)$. Since the exponential
function only affects the eigenvalues of a
matrix, the quasispecies is given by the principal
eigenvector of $\overline{\mat W}$, irrespective of the length of the
oscillation period $T$. In other words, in the absence of mutations
will the sequence $i$ with the highest average value of
$A_i(t)-D_i(t)$ take over the whole population after a suitable amount
of time, provided it existed already in the population at the
beginning of the process. By continuity, this property must extend to
very small but positive error rates $R$. So, similar to the case of
$R=0.5$, the temporal changes in the landscape loose their importance
when $R$ approaches 0.

There is, however, a caveat to the above argument. In case the largest
eigenvalue of $\overline {\mat W}$ is degenerate, temporal changes in the
landscape may continue to be of importance for $R=0$. A degeneracy of
the largest eigenvalue of $\overline {\mat W}$ is possible, because the
Frobenius-Perron theorem applies only to positive error rates. For
degenerate quasispecies, the initial condition $\vec y(t_0)$
determines the composition of the asymptotic population. In this
context, let us consider the general solution for periodic fitness
landscapes, Eq.~(\ref{eq:gen-periodic-sol}). We have
\begin{equation}
  \vec y(t) = \mat X^m(\phi)\vec y(t_0+\phi)
\end{equation}
with
\begin{equation}\label{eq:y-of-t0-p-phi}
  \vec y(t_0+\phi) = \mat Y(t_0+\phi, t_0) \vec y(t_0).
\end{equation}
So even if $\mat X$ becomes independent of $\phi$ for $R=0$, this need
not be the case for $y(t_0+\phi)$, because of
Eq.~(\ref{eq:y-of-t0-p-phi}). If the largest eigenvalue of
$\overline {\mat W}$ is degenerate, variations in $y(t_0+\phi)$ will
remain visible for arbitrarily large times $t$. Hence, we will observe
oscillations among the different quasispecies which correspond to the
largest eigenvalue. Clearly, this effect is the more pronounced the larger the
oscillation period $T$.

\subsubsection{Schematic phase diagrams}
\label{sec:schem-phase-diag}

\begin{figure}[tb]
\centerline{
 \epsfig{file={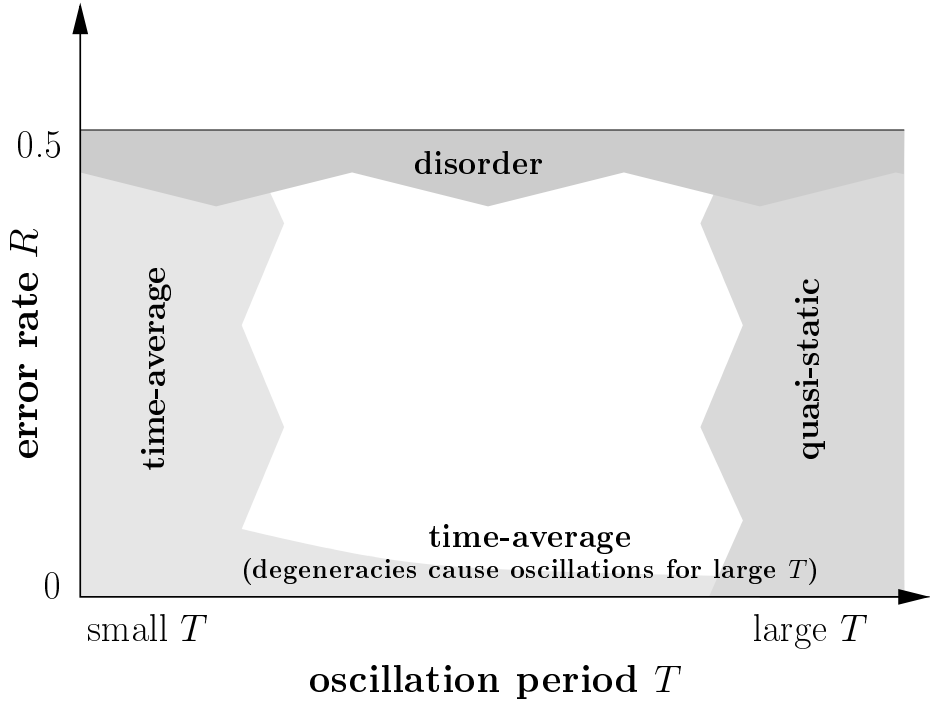}, width=.9\columnwidth}
}
\ \\[-2ex]
\caption{\label{fig:border-regions}The appearance of a periodic
 fitness landscape at the border regions of the parameter space.
}
\end{figure}

The results of the previous two subsections allow us to identify the
general properties of the quasispecies model with a periodic fitness
landscape at the borders of the parameter space. As parameters, we shall only
consider error rate $R$ and oscillation period $T$,
since all other parameters (replication rates, decay rates, details of
the matrix $\mat Q$) do not influence the above results. In
Fig.~\ref{fig:border-regions}, we have summarized our
findings. Along the abscissa runs the oscillation period. For very
fast oscillations, the evolving population sees only the time-averaged
landscape. For very slow oscillations, on the other hand, the
population is able to settle into an equilibrium much faster than the
changes in the landscape occur. Hence, the population sees a
quasistatic landscape. Along the ordinate, we have displayed the error
rate. We discarded the region above
$R=0.5$, in which anti-correlations between parent and offspring
sequences are present, as it is redundant.  For $R=0.5$, all sequences have
random offspring, and hence, all sequences replicate equally well. Therefore,
for this error rate, the landscape becomes effectively flat. On
the other side, for $R=0$, we have again the time-averaged
landscape. However, for large $T$, the fact that we see the average
landscape does not mean that the concentration variables are
asymptotically constant. Degeneracies in the largest eigenvalue may
cause a remaining time dependency due to oscillations between
superimposed quasispecies. The exact form
of these oscillations is dependent on the initial
condition $\vec y(0)$. For small $T$, the oscillations disappear, because the
ratio of newly created sequences during one oscillation period and
remaining sequences from the previous oscillation period decays with
$T$ [Eq.~(\ref{eq:high-freq-expans})].

\begin{figure}[tb]
\centerline{
 \epsfig{file={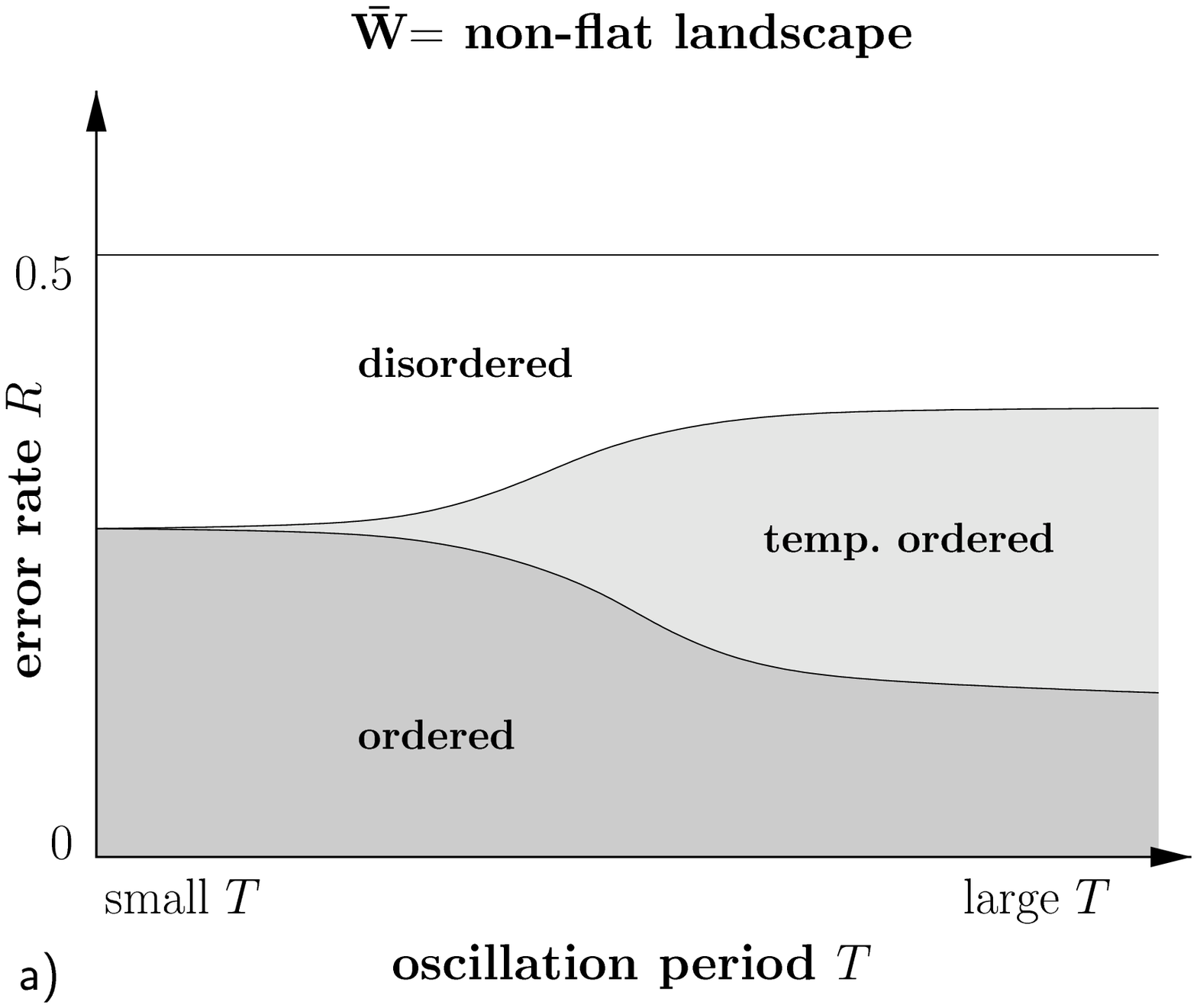}, width=.48\columnwidth}
 \epsfig{file={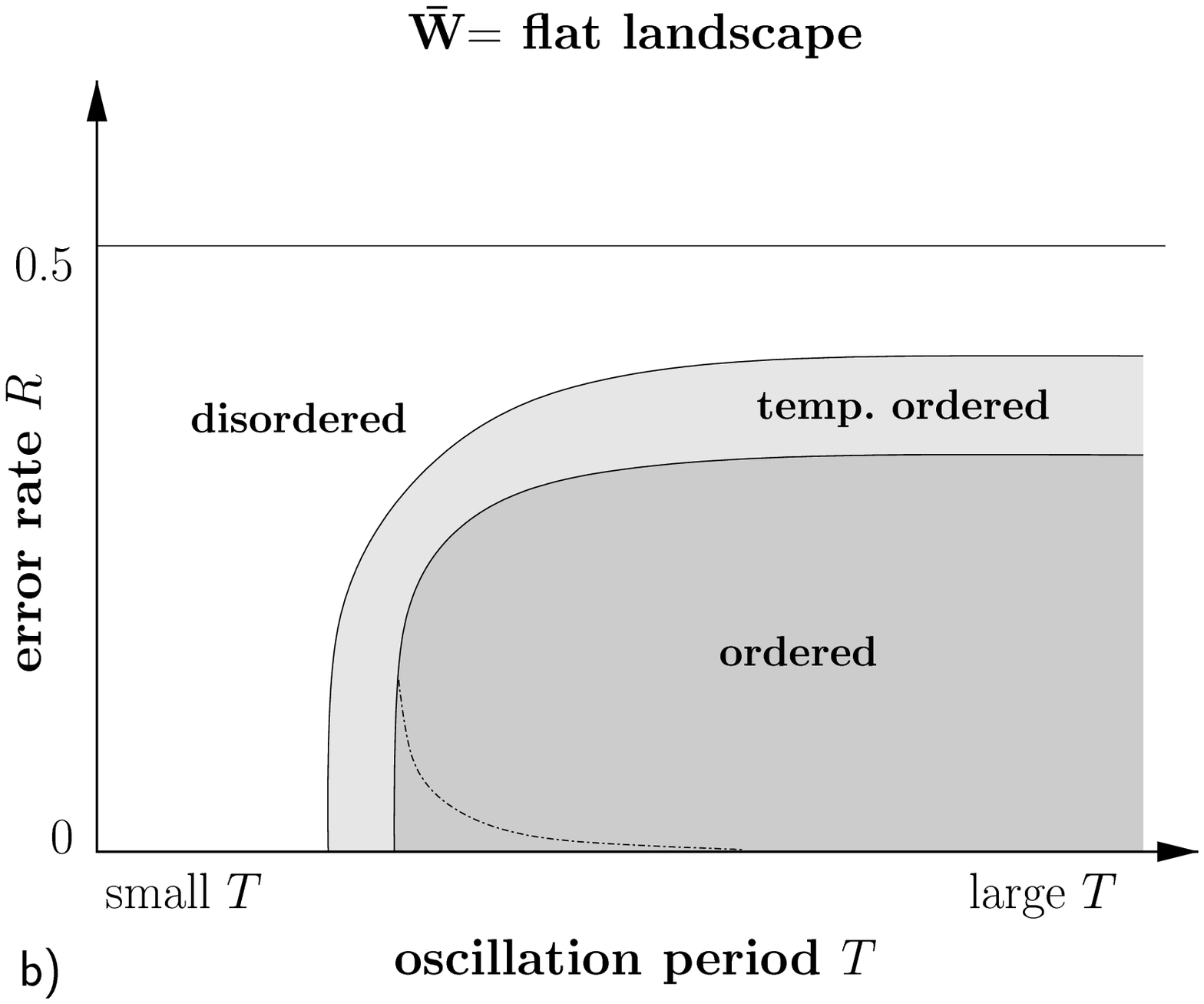}, width=.48\columnwidth}
}
\ \\[-2ex]
\caption{\label{fig:schematic-phases12}Two possible phase diagrams
 of a periodic landscape. If $\mat W(t)$ averages to a non-flat
 landscape, there will typically be a lower error threshold, below
 which we always find order, and a higher error threshold, above which
 the system is always in a disordered state. If $\mat W(t)$ averages
 to a flat landscape, however, the
 disordered phase extends to the whole range of $R$ for sufficiently
 small $T$. The dashed line inside the ordered regime in case b) is explained
 in the text.
}
\end{figure}

From the above observations, we can derive generic phase diagrams for
periodic fitness landscapes. There are two main possibilities. The
fitness landscape may average to a landscape that has a
distinct quasispecies, or it may average to a flat
landscape. These two cases are illustrated in
Fig.~\ref{fig:schematic-phases12}. Note that the diagrams are meant to
illustrate the qualitative form and position of the different
phases. In their exact appearance, they may differ substantially from
the exact phase diagram of a particular landscape.

If an averaged landscape sports a distinct quasispecies, then for
every oscillation period $T$ and every phase of the oscillation
$\phi$, we have a unique error threshold $R^*(T,\phi)$. For small $T$,
the error threshold converges towards that of the average fitness
landscape, $R^*_{\rm av}$, irrespective of the phase $\phi$. For
larger $T$, the error threshold oscillates between
$R^*_{\rm lo}=\min_\phi R^*(T,\phi)$ and $R^*_{\rm hi}=\max_\phi R^*(T,\phi)$.
In the limit of an infinitely large oscillation period, $R^*_{\rm hi}$
converges towards $R^*_{\max}$, which is the largest error threshold
of all the (static) landscapes $\mat W(\phi)$. Similarly,
$R^*_{\rm lo}$  converges towards $R^*_{\min}$ in that limit, where
$R^*_{\min}$ is accordingly defined as the smallest error threshold of
all landscapes $\mat W(\phi)$. For a fixed oscillation period $T$ and
a fixed error rate $R$ with $R^*_{\rm lo}< R < R^*_{\rm hi}$ , we have
necessarily $R > R^*(T,\phi)$  for some phases $\phi$, and $R <
R^*(T,\phi)$ for the rest of the oscillation period. As a result, a
quasispecies will form whenever $R > R^*(T,\phi)$, but it will
disappear again as soon as $R < R^*(T,\phi)$. This phenomenon has
for the first time been observed
in~\cite{WilkeRonnewinkelMartinetz99}, where the region of the
parameter space in which it can be found has been called the
\emph{temporarily ordered phase}. In this phase, whether we observe order or
disorder depends on the particular moment in time at which we study
the system. Correspondingly, we will call a phase
``ordered'' only if order can be seen for the whole oscillation
period, and we will call a phase ``disordered'' if during the whole
oscillation period no order can be seen. The relationship between the
ordered phase, the disordered phase, and the temporarily ordered phase
for the first type of landscapes is displayed in
Fig.~\ref{fig:schematic-phases12}a. Compare also the phase diagram of
the oscillating Swetina-Schuster landscape in Fig.~\ref{fig:phases-l10n200}.

In a landscape that averages to a flat one, on the other hand, the
disordered phase must extend over the whole range of $R$ for
sufficiently small $T$, and order can be observed only above a certain
$T_{\min}$. The behavior of the system for small $R$ above $T_{\min}$ cannot
in general be predicted solely from the knowledge of
Fig.~\ref{fig:border-regions}. For a landscape with a flat average, the
eigenvalues of the monodromy matrix are degenerate for $R=0$. Therefore, in
the limit $R\rightarrow 0$, the Perron eigenvector can, in principle, converge
to any superposition of the eigenvectors of $\mat X(R=0)$. This situation is
visualized in Fig.~\ref{fig:schematic-phases12}b. If the limit corresponds to
a non-homogeneous sequence distribution, the ordered phase extends to
arbitrarily small $R$ [indicated by the solid line in
Fig.~\ref{fig:schematic-phases12}b]. If, on the other hand, the limit
would correspond to a homogeneous sequence distribution, we might find a lower
error
threshold below which the system would again be in disorder [this is indicated
by the dashed line in Fig.~\ref{fig:schematic-phases12}b]. Since for longer
oscillation periods, the oscillations in the degenerate quasispecies at $R=0$
become important, order would be observed for much smaller $R$ with increasing
$T$. Hence, the lower disordered phase would fade out for
$T\rightarrow\infty$. A study investigating under which situations the ordered
phase extends to arbitrarily small $R$ will be presented
elsewhere. As a preliminary result, we can state that the limit $R\rightarrow0$
does in general not lead to a homogeneous sequence distribution.

\subsection{Discrete approximation}
\label{sec:discrete-approx}

The differential equation formalism we have used so far allows for an
elegant
discussion of the system's general properties. However, if we want to
obtain numerical solutions, this formalism is not very helpful,
because
we do not have a general expression for the fundamental matrix $\mat
Y(t, t_0)$ from Eq.~(\ref{eq:lin-solution}), nor for the monodromy
matrix $\mat X(t_0)$ from Eq.~(\ref{eq:def-monodromy}). Therefore, for
a numerical treatment we need to move over to the discretized
quasispecies equation,
\begin{equation}\label{eq:basic-discrete}
  \vec y(t+\Delta t) = [\Delta t \mat W(t) +\Idmat] \vec y(t)\,.
\end{equation}
In the case of constant $\mat W$, the quasispecies obtained
from this equation is identical to that of
Eq.~(\ref{eq:basic-linearized-eq}), and it is also identical to that of
equation
\begin{equation}\label{eq:standard-discrete-eq}
  \vec y(t+1) = \mat W \vec y(t)\,.
\end{equation}
Equation~(\ref{eq:standard-discrete-eq}) has been studied by
Demetrius \etal~\cite{Demetriusetal85}, and has been employed by
Leuth\"ausser \cite{Leuthaeusser85,Leuthaeusser87}
for her mapping of the quasispecies model onto the Ising model. In the
general time-dependent case, however, the additional factor $\Delta t$
and the identity matrix $\Idmat$ of Eq.~(\ref{eq:basic-discrete}) are
important, and cannot be left
out. The analogue of the fundamental matrix for
Eq.~(\ref{eq:basic-discrete})
reads
\begin{equation}
  \mat Y(t_0+k\Delta t, t_0)={\mathcal{T}}\left\{
     \prod_{\nu=0}^{k-1}[\Delta t \mat W(t_0+\nu\Delta
     t)+\Idmat]\right\}\,,
\end{equation}
where ${\mathcal{T}}\{\cdot\}$ indicates that the matrix product has to be
evaluated with the proper time ordering~\cite{Wilke99}. Similarly, the
analogue of
the monodromy matrix becomes
\begin{align}\label{eq:discreteX}
  \mat X(t_0)&=\mat Y(t_0+T, t_0)\notag\\
  & = {\mathcal{T}}\left\{
     \prod_{\nu=0}^{n-1}[\Delta t \mat W(t_0+\nu\Delta
     t)+\Idmat]\right\}\,,
\end{align}
where we have assumed that $T$ is an integral multiple of $\Delta t$,
and where $n=T/\Delta t$. The influence of the size of $\Delta t$
on the quality of the approximation has been investigated in~\cite{Wilke99}.
A more in-depths discussion of the relationship between the continuous
and the discrete quasispecies model can also be found in~\cite{BaakeGabriel99}.

\subsection{Example landscapes}
\label{sec:example-landscapes}

For the rest of this section, we are going to study several
example landscapes, in order to illustrate the implications of our
general theory. In all cases considered, we represent the molecules as
bitstrings of fixed length $l$. Moreover, we assume that a single bit
is copied erroneously with rate $R$, irrespective of the bit's type
and of its position in the string.

\subsubsection{One oscillating peak}

In previous work on the quasispecies model with periodic fitness
landscapes~\cite{Wilke99,WilkeRonnewinkelMartinetz99}, most emphasis
has been laid on landscapes with a single oscillating sharp peak. As a
generalization of the work of Swetina and
Schuster~\cite{SwetinaSchuster82}, the master sequence has been given
a replication rate $A_0(t)\gg A$, where $A$ is the replication rate of
all other sequences. The replication rate $A_0(t)$ has been expressed as
\begin{equation}
  A_0(t)=A_{0, \rm stat}\exp[\epsilon f(t)]\,,
\end{equation}
with a $T$-periodic function $f(t)$. The generic example for that function is
$f(t)=\sin(\omega t)$, leading to
\begin{equation}\label{eq:single-osc-peak}
  A_0(t)=A_{0, \rm stat}\exp[\epsilon \sin(\omega t)]\,,
\end{equation}
The parameter $\epsilon$ allows a
smooth crossover from a static landscape to one with considerable
dynamics, and the exponential assures that $A_0(t)$ is always
positive.

In~\cite{Wilke99,WilkeRonnewinkelMartinetz99}
it has been found that the behavior at the border regions of the
parameter space is indeed as it is depicted in
Fig.~\ref{fig:border-regions}, and that a phase diagram of the form of
Fig.~\ref{fig:schematic-phases12}a correctly describes the
relationship of order and disorder in an oscillating Swetina-Schuster
landscape. Here, we put less emphasis on the numerical
simulations that lead to these conclusions, but instead show that
the phase borders in such a phase diagram can, for an oscillating
Swetina-Schuster landscape, be calculated approximately.

For static landscapes with a single peak, the assumption of a
vanishing mutational backflow into the master sequence allows to
derive an approximate expression for the error
threshold~\cite{MaynardSmith83,Eigenetal88,Eigenetal89}. A similar
formula can be developed to calculate the
error threshold as a function of time in a landscape with a single
oscillating peak.
But before we turn towards the dynamic landscape, we shall rederive
the expression for
the master sequence's concentration $x_0$ in a static landscape, based on
neglecting mutational backflow. The expression we shall find is
slightly more general than the one that was previously given, and it
will be of use for the periodic fitness landscape as well.

The 0th component of the quasispecies
equation~(\ref{eq:basic-quasispecies}) becomes, after neglecting the
mutational backflow,
\begin{equation}\label{eq:no-mut-backflow}
 \dot x_0(t) = W_{00}x_0(t) - \overline E(t)x_0(t)\,.
\end{equation}
The average excess production $\overline E(t)$ can be expressed in terms of
$\vec x(t)$ and $\mat W$ as
\begin{equation}\label{eq:no-backflow-barE1}
  \overline E(t) = \sum_{i,j} W_{ij}x_j(t)\,.
\end{equation}
With that expression, the solution of Eq.~(\ref{eq:no-mut-backflow})
requires the knowledge of the stationary mutant concentrations $x_j$,
which are usually unknown. To circumvent this problem, we make the
somewhat extreme assumption that all mutant concentrations are
equal. Although this assumption, which is equivalent to the assumption
of equal excess productions $E_i$ in the usual calculation without
mutational backflow, will generally not be true, it works fine for
Swetina-Schuster type landscapes. With
this additional assumption, Eq.~(\ref{eq:no-backflow-barE1}) becomes
\begin{equation}
  \overline E(t)=\sum_i\left [\sum_{j>0}W_{ij}\frac{1-x_0(t)}{N-1}
    + W_{i0}x_0(t)\right]\,,
\end{equation}
where $N$ is the number of different sequences in the system. When we
insert this into Eq.~(\ref{eq:no-mut-backflow}) and solve for the
steady state, we find
\begin{equation}\label{eq:no-backflow-solution}
  x_0 = \frac{W_{00} -\frac{1}{N-1} \sum_i\sum_{j>0}W_{ij}}
     {\sum_i W_{i0} - \frac{1}{N-1}\sum_i\sum_{j>0}W_{ij}}\,.
\end{equation}
The expressions involving sums over matrix elements in
Eq.~(\ref{eq:no-backflow-solution}) can be identified with the excess
production of the master,
\begin{equation}
  E_0=\sum_iW_{i0}
\end{equation}
and with the average excess production without the master,
\begin{equation}
  \overline E_{-0}= \frac{1}{N-1} \sum_i\sum_{j>0} W_{ij}\,,
\end{equation}
if $\mat W$ has the standard form $\mat Q\mat A - \mat D$. Therefore,
Eq.~(\ref{eq:no-backflow-solution}) corresponds to the often quoted
result
\begin{equation}
  x_0=\frac{W_{00} - \overline E_{-0}}{E_0 -\overline E_{-0}}\,.
\end{equation}
However, Eq.~(\ref{eq:no-backflow-solution}) is more general in that
it can be used even if $\mat W$ is not given as $\mat Q\mat A-\mat D$.

The idea here is to insert the monodromy matrix into
Eq.~(\ref{eq:no-backflow-solution}) in order to obtain an
approximation for $x_0$ in the case of periodic landscapes. But under what
circumstances can we expect this to work? After all,
Eq.~(\ref{eq:no-backflow-solution}) has been derived from an equation
with continuous time, Eq.~(\ref{eq:no-mut-backflow}), whereas the
monodromy matrix advances the system in discrete time steps, as can
bee seen in Eq.~(\ref{eq:basic-mon-solution}). The important point is
here that we are only interested in the asymptotic state, which is
given by the normalized Perron vector of the monodromy matrix, whether we use
discrete or continuous time. Therefore, we are free to calculate the
asymptotic state in a periodic landscape for a given phase $\phi$ from
\begin{equation}\label{eq:mon-to-conttime1}
  \dot{\vec y}(t) = \mat X(\phi)\vec y(t)\,,
\end{equation}
even if this equation does not have a direct physical meaning for
finite times. The asymptotic molecular concentrations are then given
by the limit $t\rightarrow\infty$ of
\begin{equation}\label{eq:mon-to-conttime2}
  \vec x(t) = \frac{\vec y(t)}{\vec e\cdot \vec y(t)}\,.
\end{equation}
From differentiating Eq.~(\ref{eq:mon-to-conttime2}) and inserting
Eq.~(\ref{eq:mon-to-conttime1}), we obtain
\begin{equation}
  \dot{\vec x}(t)=\mat X(\phi)\vec x(t)-\vec x(t)
     \big( \vec e\cdot[\mat X(\phi)\vec x(t)]\big)\,.
\end{equation}
When we neglect the backflow onto the master sequence, the 0th component
of that equation becomes identical to Eqs.~(\ref{eq:no-mut-backflow})
and (\ref{eq:no-backflow-barE1}), but with the matrix $\mat X(\phi)$
instead of $\mat W$. This shows that we may indeed use
Eq.~(\ref{eq:no-backflow-solution}) as an approximation for the
asymptotic concentration of $x_0$. Of course, since we have neglected
mutational backflow, this approximation works only for landscapes in
which a single sequence has a significant advantage over all
others. But this restriction does equally apply to the static
case. Numerically, we have found that
Eq.~(\ref{eq:no-backflow-solution}) works well for a single oscillating
peak, and that it breaks down in other cases as expected.

\begin{figure}[t]
\centerline{
 \epsfig{file={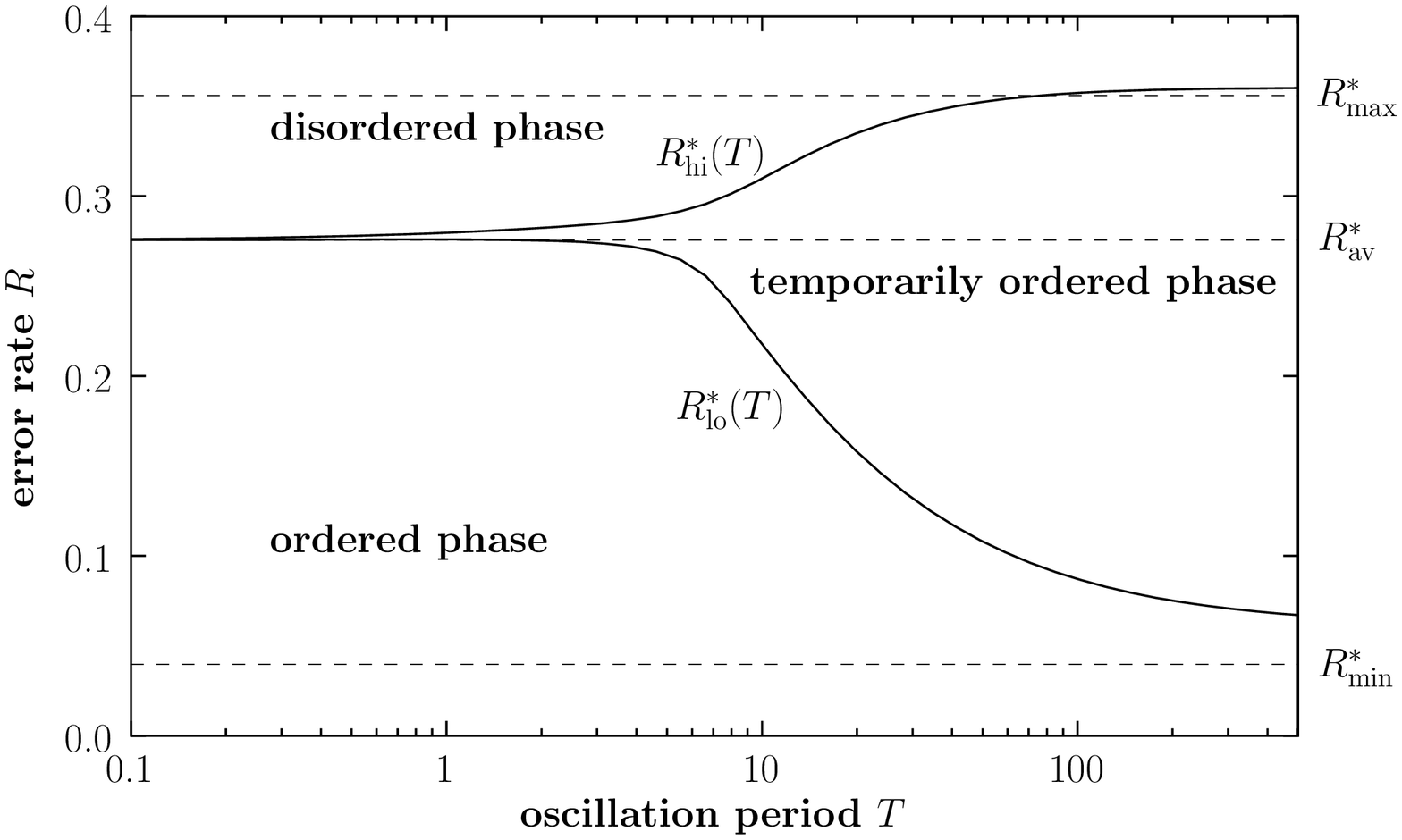}, width=0.9\columnwidth}
}
\ \\[-2ex]
\caption{\label{fig:phases-l10n200}The phase diagram of an oscillating
 Swetina-Schuster landscape $[A_0(t)=e^{2.4}\exp(2\sin \omega t)]$,
 obtained numerically from Eq.~(\ref{eq:no-backflow-solution}).
}
\end{figure}

 With
the aid of Eq.~(\ref{eq:no-backflow-solution}), we are now in the
position to calculate the phase diagram of the
oscillating Swetina-Schuster landscape. When we
insert the monodromy matrix $\mat X(\phi)$ into
Eq.~(\ref{eq:no-backflow-solution}), we are able to obtain
(numerically) the error rate at which $x_0$ vanishes, $R^*(T,\phi)$.
From that expression, we can calculate $R^*_{\rm lo}$ and $R^*_{\rm hi}$.
 The results of the
corresponding, numerically extensive calculations are shown in
Fig.~\ref{fig:phases-l10n200}, together with $R^*_{\rm av}$, $R^*_{\max}$, and
$R^*_{\min}$, which have also been determined from
Eq.~(\ref{eq:no-backflow-solution}).

We find that both $R^*_{\rm lo}$ and $R^*_{\rm hi}$ approach
$R^*_{\rm av}$ for $T\rightarrow 0$, as predicted by our general
theory. For $T\rightarrow\infty$, $R^*_{\rm hi}$ grows quickly to the
level of $R^*_{\max}$, but a slight discrepancy between the two values
remains. This is due to the complexity of the numerical
calculations involved for large $T$. We can only approximate the
monodromy matrix by means of Eq.~(\ref{eq:discreteX}), and we need
ever more factors $\Delta t \mat W(t_0+\nu\Delta t)+\Idmat$ for large
$T$. The discrepancy between $R^*_{\rm lo}$ and $R^*_{\min}$, on the
other hand, has a different origin. The main cause here is the fact
that the relaxation into equilibrium is generally slower for smaller
error rates. Therefore, $R^*_{\rm lo}$ needs a much larger $T$ to
reach $R^*_{\min}$ than it is the case with $R^*_{\rm hi}$ and
$R^*_{\max}$.

Very recently, Nilsson and Snoad~\cite{NilssonSnoad2000a} have
demonstrated that the method of neglecting back mutations described above can
be used to calculate an approximate analytic solution for the oscillating
peak. They exploit the fact that if all replication rates except $A_0(t)$
are equal to 1, Eq.~(\ref{eq:no-mut-backflow}) becomes
\begin{equation}\label{eq:NS-osc-peak}
  \dot x_0(t) = A_0(t) x_0(t) - \bar E(t)x_0(t)\,,
\end{equation}
with
\begin{equation}
  \bar E(t) = [ A_0(t) - 1] x_0(t) + 1
\end{equation}
and $Q=(1-R)^l$. Eq.~(\ref{eq:NS-osc-peak}) can then be solved exactly by
introducing a nonlinear transformation
\begin{equation}
  x_0(t) \rightarrow y(t) =\frac{Q-x_0(t)}{[1-Q]x_0(t)},
\end{equation}
which linearizes the equation. For large $t$, Nilsson and Snoad obtain
\begin{equation}\label{eq:NS-osc-peak-sol}
  x_0(t) = Q\left[1+(1-Q)\int_0^t e^{-\int_s^t[QA_0(u)-1]\,du} ds\right]^{-1}\,.
\end{equation}
Interestingly, this expression holds for arbitrary $A_0(t)$, and not only for
periodic ones.

From our discussion in Sec.~\ref{sec:diff-equ-formalism}, we know that the
master sequence's concentration will, for fast changes, settle to the value
corresponding to the average replication rate, and for very slow changes it
will be virtually in equilibrium with the current replication rate. In the
intermediate regime, we expect the concentration to lag somewhat behind the
replication rate. This behavior can be seen as a low-pass filtering of the
environmental changes, performed by the evolving population. The idea to look
at time-dependent fitness landscapes from that perspective was first
introduced by Hirst~\cite{Hirst97a,HirstRowe99}, who reevaluated similar
observations made in population genetic models (without explicit
landscape)~\cite{SasakiIwasa87,IshiiMatsudaIwasaSasaki89,Charlesworth93,LandeShannon96}.
For the quasispecies model, phase shift and amplitude of the response to a
sinusoidal replication rate have been determined computationally
in~\cite{Wilke99}, with the result that these curves do indeed have the
appropriate form for a low pass filter. Based on
Eq.~(\ref{eq:NS-osc-peak-sol}), an analytic expression for amplitude and phase
shift has been given in~\cite{NilssonSnoad2000a}.

\subsubsection{Validity of the expansion in $T$}
\label{sec:validity-of-expansion}

In the previous section, we have calculated the phase diagram in a landscape
with a single oscillating peak. In this section, we are going to asses the
validity of the monodromy expansion in $T$ in a similar landscape. Since the
integrals involved become very unpleasant if we choose the master sequence's
replication rate to be proportional to $\exp[\sin(\omega t)]$, we study here
the related landscape
\begin{subequations}
\begin{align}\label{eq:single-osc-peak-expans}
   A_0(t)&= A_{0, \rm stat}[1+\epsilon \sin(\omega t)]\,,\\
   A_i(t)&= 1\quad \mbox{for $i>0$.}
\end{align}
\end{subequations}
This landscape follows from expanding Eq.~(\ref{eq:single-osc-peak}) to first
order in $\epsilon$. Eq.~(\ref{eq:w-av-arbitrary-order}) can be evaluated
analytically to arbitrary order for that landscape. In
Appendix~\ref{app:high-frequ-ex}, we have carried out the corresponding
calculations to 2nd order in $T$.

\begin{figure}[tb]
\centerline{
 \epsfig{file={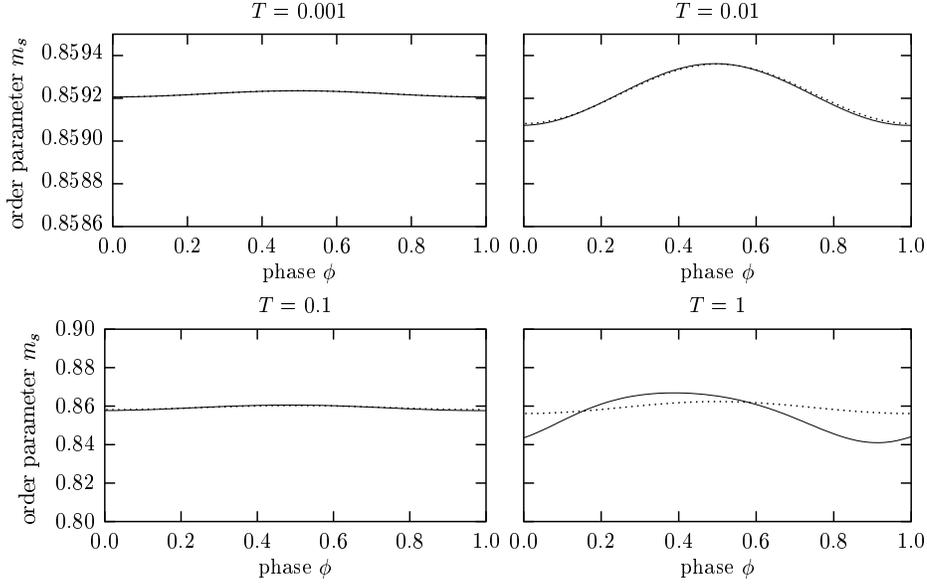}, width=0.9\columnwidth}
}
\ \\[-2ex]
\caption{\label{fig:app-vs-num}Order parameter $m_s$ for a landscape with
 a single oscillating peak, Eq.~(\ref{eq:single-osc-peak-expans}).
 The error rate is $R=0.06$. Solid lines stem from
 the discrete approximation with $n=100$,
 the dotted lines stem from the expansion in terms of $T$,
 Eq.~(\ref{eq:X-Neumann-exp}), evaluated up to second order. Clearly,
 the expansion Eq.~(\ref{eq:X-Neumann-exp}) is only of use for
 relatively short oscillation periods.
}
\end{figure}

In Fig.~\ref{fig:app-vs-num}, we display the order parameter
$m_s$ obtained from the expansion of $\mat X$ in terms of $T$ and from
the discrete approximation of $\mat X$ as a function of the phase
$\phi$ for four different oscillation periods $T$. The results from the two
different methods agree well for $T\leq 0.1$, but disagree for
larger $T$. The disagreement results from a breakdown of the expansion in $T$
when $T$ becomes large. Note that this breakdown occurs in the vicinity of the
radius of convergence that we have estimated in
Eq.~(\ref{eq:radius-of-convergence}). For the particular landscape of
Fig.~\ref{fig:app-vs-num}, the estimate guarantees convergence for oscillation
periods below approx.\ 0.1.

\subsubsection{Two oscillating peaks}

A single oscillating peak provides some initial insights into dynamic
fitness landscapes. It
is more interesting, however, to study situations in which several
sequences obtain the highest replication rate in different phases of the
oscillation period. The simplest such case is a landscape in which two
sequences in turn become the master sequence. Here, we will assume
that the two are located at opposite corners of the boolean hypercube,
i.e., that they are given by a sequence and its
inverse. In that way, it is possible to group sequences into
error classes according to their Hamming distance to one of the two possible
master sequences. As an example, we are going to study a landscape
with the replication coefficients
\begin{subequations}\label{eq:two-peaks-fitness}
\begin{align}
  A_0(t) &= A_{0,\rm stat}\exp(\epsilon\sin \omega t)\,,\\
  A_l(t) &= A_{0,\rm stat}\exp(-\epsilon\sin \omega t)\,,\\
  A_i(t) &= 1 \quad \mbox{for $0<i<l$.}
\end{align}
\end{subequations}
The subscripts in the replication coefficients stand for the Hamming
distance to the sequence $000\cdots0$.

For single peak landscapes, it is instructive to characterize the
state of the system at time $t$ by the value of the order parameter $m_s(t)$
[Eq.~(\ref{eq:def-order-parameter})].
In principle,
$m_s(t)$ can also be used for a landscape
with two alternating master sequences if they are each other's
inverse. In that case, the Hamming distance has to be
measured with respect to one of the two master sequences. If the
population consists only of sequences of the other type of master
sequence, we have $m_s(t)=-1$. However, there is a small problem with
degenerate landscapes in which the two peaks have the same
replication rate. In such landscapes, the sequence distribution
becomes symmetric with respect to the two peaks, i.e., $x_0=x_l$,
$x_1=x_{l-1}$, and so on. Then, $m_s(t)$
becomes zero because of this symmetry, although the population may be
in an ordered state. To distinguish between the case of true
disorder and the case of an ordered, but symmetrical population, we
introduce the additional order parameters
\begin{equation}\label{eq:order-param-mplus}
  m_s^+(t) = \frac{1}{l} \sum_{i=0}^{\lfloor(l-1)/2\rfloor}
  x_i(t)[l-2i]\,,
\end{equation}
and
\begin{equation}
  m_s^-(t) = \frac{1}{l} \sum_{i=l-\lfloor(l-1)/2\rfloor}^l
  x_i(t)[l-2i]\,.
\end{equation}
Here, $\lfloor x\rfloor$ stands for the largest integer smaller than
or equal to $x$.

The quantity $m_s^+(t)$ is always positive, $m_s^-(t)$ is always
negative, and furthermore, we have
\begin{equation}
  m_s(t) = m_s^+(t) + m_s^-(t)\,.
\end{equation}
If the population is uniformly distributed over the whole sequence
space, we have
\begin{equation}\label{eq:mspm-disordered}
  m_s^+(t) = - m_s^-(t)=\frac{1}{l2^l}
  \sum_{i=0}^{\lfloor(l-1)/2\rfloor}
  \binom{l}{i}(l-2i)\,.
\end{equation}
This expression tends to 0 for $l\rightarrow \infty$. If, on the other
hand, only the two peaks are populated, each with half of the total
population, we find
\begin{equation}
  m_s^+(t) = - m_s^-(t)=\frac{1}{2}\,.
\end{equation}
In the case that either  $m_s^+(t)$ or $m_s^-(t)$ vanish, the
population is centered about the respective other peak.

\begin{figure}[t]
\centerline{
 \epsfig{file={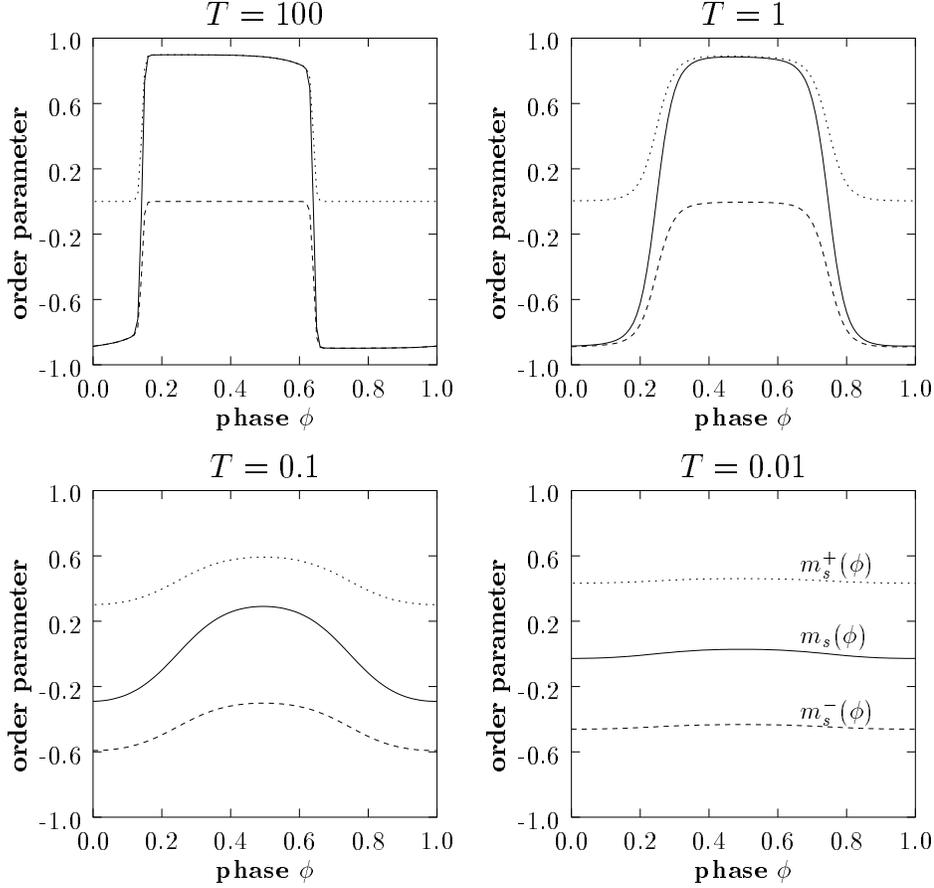}, width=0.9\columnwidth}
}
\ \\[-2ex]
\caption{\label{fig:2masterTvarR0.05}Order parameters $m_s(t)$,
 $m_s^+(t)$, $m_s^-(t)$ as a function of the oscillation phase
 $\phi=(t\mod T)/T$ in a landscape with two alternating peaks. The upper dashed
 line represents $m_s^+(t)$, the lower dashed
 line represents $m_s^-(t)$, and the solid line represents
 $m_s(t)$. The sequence length is $l=10$, and we have used $R=0.05$
 and $n=T/\Delta t=100$ in all four examples. The parameters of the
 fitness landscape are $A_{0, \rm stat}=e^{2.4}$, $\epsilon = 2$.
}
\end{figure}

In the following, when it is important to distinguish between true
disordered populations and symmetric populations, we will use
$m_s^+(t)$ and $m_s^-(t)$. When the situation is non-ambiguous, we
will use $m_s(t)$ alone, in order to improve the clarity of our figures.

In Fig.~\ref{fig:2masterTvarR0.05}, we have displayed $m_s^+(t)$,
$m_s^-(t)$ and $m_s(t)$ for the quasispecies in a fitness
landscape of the type defined in Eq.~(\ref{eq:two-peaks-fitness}). For
a large oscillation period, $T=100$, the quasispecies is at every
point in time clearly centered around a single peak. The switch
from one peak to the other happens very fast. When the landscape
oscillates with a higher frequency, the transition time becomes a
larger fraction of the total oscillation period. This makes the
transition from one peak to the other appear softer in the graphs for
smaller oscillation periods.
For extremely small oscillations, the system perceives the
average fitness landscape, which is a degenerate landscape with two
peaks of equal height.  As noted above, the quasispecies becomes
symmetric in such a landscape. In the lower right of
Fig.~\ref{fig:2masterTvarR0.05}, for $T=0.01$, we can identify this
limiting behavior. Both $m_s^+(t)$ and $m_s^-(t)$ are nearly constant
over the whole oscillation period with an absolute value close to
0.5. The deviation from 0.5 stems from the finite value of the error
rate, $R=0.05$ in this example. We observe further that $m_s(t)$ lies
very close to zero, thus wrongly indicating a disordered state. Note
that the absolute value of $m_s^\pm(t)$ for a uniformly spread
population lies, for the parameters of this example, at 0.12 according to
Eq.~(\ref{eq:mspm-disordered}).

Observations from the landscape with two oscillating peaks have to
be interpreted in the light of results of Schuster and Swetina on
static landscapes with two peaks~\cite{SchusterSwetina88}, who have found
that if the peaks are far away in Hamming distance (which is the case
here), a quasispecies is generally unable to occupy both peaks at the
same time, unless
they are of exactly the same height\footnote{This is true for infinite populations only. For
  finite populations, one of the two peaks will be lost eventually due
  to sampling fluctuations.}.
For two peaks with different heights, the quasispecies will for small
$R$ generally form around the higher peak. For larger $R$, however, the
quasispecies moves to the lower peak if it has a higher
mutational backflow from mutants, which is the case, for example, if
the second peak is broader than the first one. The transition from the
higher peak to the lower one with increasing $R$ is very sharp, and
can be considered a phase transition. In a dynamic landscape with
relatively slow changes, the quasispecies therefore switches the peak
quickly when the higher peak becomes the lower one and vice versa.

\begin{figure}[t]
\centerline{
 \epsfig{file={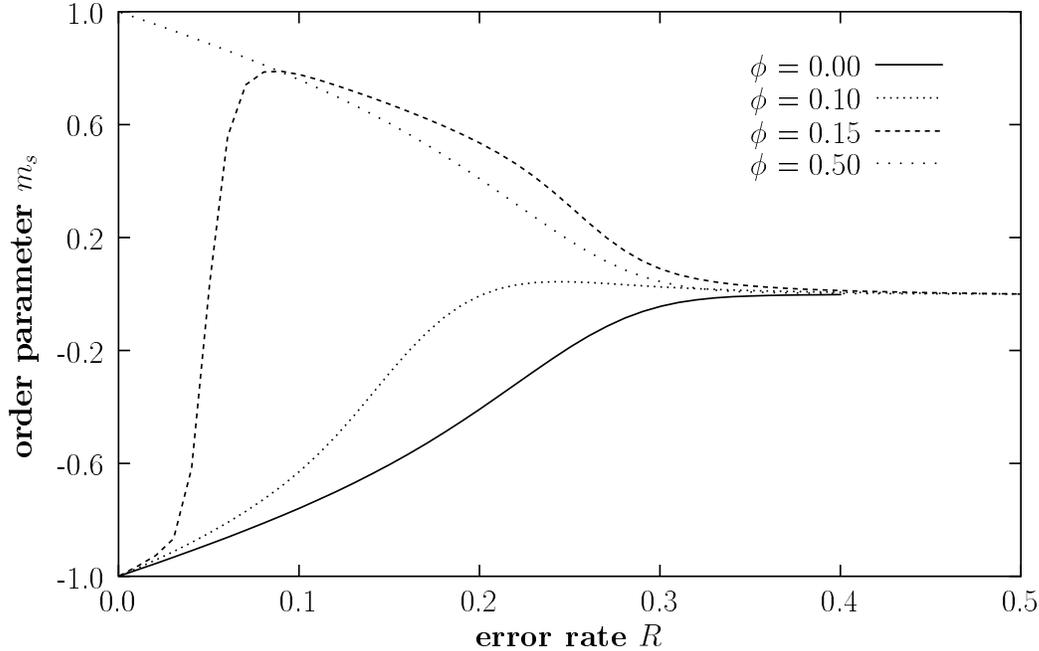}, width=1\columnwidth}
}
\ \\[-2ex]
\caption{\label{fig:2masterT100Rvar}Order parameter $m_s$ as a
 function of the error rate $R$ for various oscillation phases
 $\phi=(t\mod T)/T$. The fitness landscape is identical to the one of
 Fig.~\ref{fig:2masterTvarR0.05}, and the oscillation period is
 $T=100$. Note that for $\phi=0.10$, the error threshold seems to have
 moved towards lower $R$, which is not the case. Instead,
 the population is symmetric, as would be revealed by graphing $m_s^+$ or
 $m_s^-$.
}
\end{figure}

The exact time at which the switch occurs depends of course on the
error rate. The lower the error rate, the longer the population
remains centered around the previously higher peak until it actually
moves on to the new higher peak. Therefore, if we observe the system
at a fixed phase and change the error rate, the
quasispecies undergoes, for certain phases $\phi$, a transition
similar to the one found in~\cite{SchusterSwetina88} for static
landscapes. This is illustrated in Fig.~\ref{fig:2masterT100Rvar},
where we display the order parameter $m_s$ as a function of the error
rate $R$. At the beginning of the oscillation period the
quasispecies is, for all error rates $R$ below the error threshold,
dominated by the peak corresponding to $m_s=-1$. This must be the
case, as the replication coefficients of the two peaks intersect at
$\phi=0$, so up to this point the quasispecies has not had a chance to
build up around the other peak. For phases shortly after $\phi=0$, the
quasispecies gains weight around the other peak, starting from the
error threshold on downwards. For $\phi=0.15$, we observe
a relatively
sharp transition from the peak corresponding to $m_s=-1$ to the peak
corresponding to $m_s+1$ at $R\approx0.05$. The transition
then moves quickly towards $R=0$, until the peak corresponding to
$m_s=1$ dominates the quasispecies for all $R$. For $\phi=0.5$, the
replication coefficients intersect again, and the quasispecies is
exactly the inverse of the one for $\phi=0$.

\section{Aperiodic or stochastic fitness landscapes}
\label{sec:aperiodic-landscapes}

As we have seen, periodic fitness landscapes can be treated rather
elegantly. We have been able to define a
meaningful quasispecies, as well as to determine the general dynamics in the
border regions of the parameter
space. It would be desirable to obtain similar results for arbitrary
dynamic landscapes. After all, an aperiodic or stochastic change is
much more realistic than an exactly periodic
change. However, the definition of a time-dependent
quasispecies is tightly connected to periodic fitness landscapes. For
arbitrary changes, the concept of an asymptotic state ceases to be meaningful.
Regardless of that, we can derive some results for the border
regions of the parameter space. In
Section~\ref{sec:diff-equ-formalism}, we derived the formal solution
to Eq.~(\ref{eq:basic-linearized-eq}),
\begin{equation}
  \vec y(t_0+\tau) =  \sum_{k=0}^\infty \tau^k \overline{\mat W}_k(t_0,\tau)\vec y (t_0)\,.
\end{equation}
This can be expanded to first order in $\tau$ as
\begin{equation}\label{eq:stoch-formal-sol}
  \vec y(t_0+\tau) = \vec y(t_0) + \tau \overline{\mat W}_1(t_0,\tau)\vec y (t_0)\,.
\end{equation}
Obviously, the composition of the sequence distribution changes
very little over the interval $[t_0, t_0+\tau]$ if the condition
\begin{equation}
  \tau\norm{\overline{\mat W}_1(t_0,\tau)}_1 \ll 1
\end{equation}
is satisfied. This observation allows us to establish a general result
for rapidly changing fitness landscapes. If the landscape changes in
such a way that for every interval of length $\tau$ beginning at time $t_0$,
the average
\begin{equation}
   \overline{\mat W}_1(t_0,\tau) = \frac{1}{\tau}\int_0^{\tau} \mat W(t_0+\tau_1)
     d\tau_1
\end{equation}
is approximately the same for every $t_0$, and the condition
$\norm{\smash{\overline{\mat W}_1(t_0,\tau)}} \ll 1/\tau$ holds, then the system
develops a quasispecies given by the normalized principal eigenvector
of the average matrix $\overline{\mat W}_1(t_0,\tau)$. With
``approximately the same'' we mean that for two times $t_0$ and $t_1$,
the components of the averaged matrices satisfy
\begin{equation}
  \left| \Big(\overline{\mat W}_1(t_0,\tau)\Big)_{ij} -
    \Big(\overline{\mat W}_1(t_1,\tau)\Big)_{ij} \right| < \epsilon
  \quad \mbox{for all $i$, $j$, $t_0$, $t_1$,}
\end{equation}
with a suitably small $\epsilon$. In other words, if the fitness
landscape changes very fast but in stationary way, then the evolving
population sees only the time-averaged fitness landscape.

\begin{figure}[tb]
\centerline{
 \epsfig{file={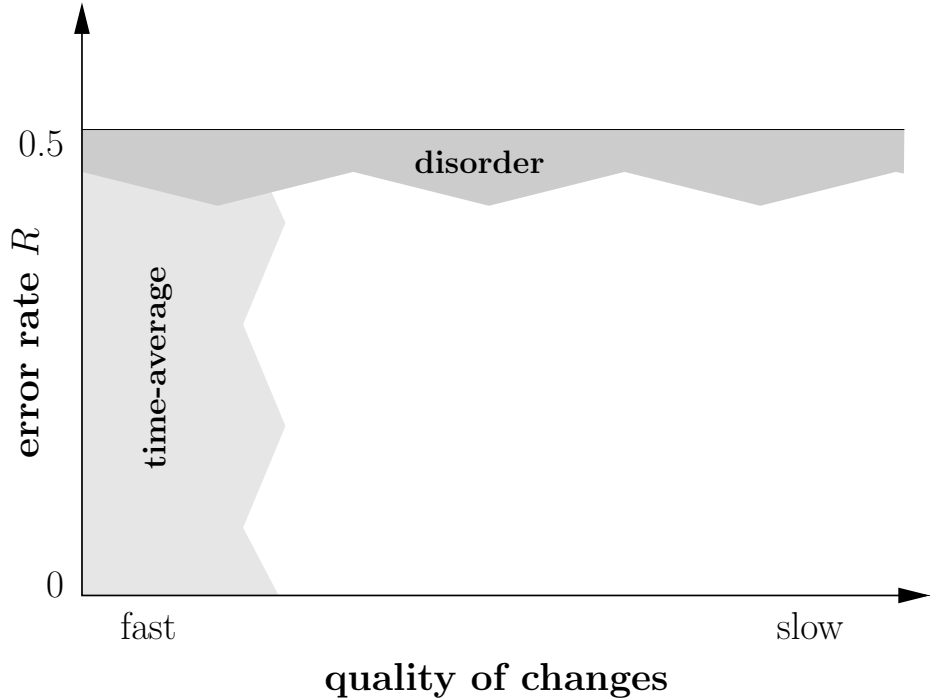}, width=.9\columnwidth}
}
\ \\[-2ex]
\caption{\label{fig:border-regions-stoch}The appearance of a stochastic
 fitness landscape at the border regions of the parameter space.
}
\end{figure}

For the special case of $R=0$ we can, as in
Eq.~(\ref{eq:sol-basic-R0-eq}), write the solution for the
quasispecies equation as
\begin{equation}
  \vec y(t) = \exp\left(\int_{t_0}^t [\mat A(t')-\mat D(t')]dt'\right)
               \vec y(t_0)\,.
\end{equation}
Unlike in the case of a periodic landscape, however, this does not
tell us the general behavior at $R=0$, apart from the fact that for
fast changes, the system experiences the average fitness landscape. For
stochastic landscapes with long time correlations, it is hard to make general
statements. The reason for
this is that from long time correlations, we cannot generally deduce
that the system must be in a quasistatic state. In a landscape that is
constant most of the time, but displays intermittent sudden changes, the
system can be expected to be mostly in the quasistatic regime. However, one
can easily construct landscapes that are in constant flux,
and still display long time correlations. Hence, there exists no
direct equivalent to the case of periodic
fitness landscapes with large oscillation period for general stochastic
landscapes. Nevertheless, we
can draw a diagram similar to Fig.~\ref{fig:border-regions}, where for the
abscissa, we use the qualitative description ``slow'' and ``fast''
changes. Under ``fast'', we subsume everything that satisfies the
above stated conditions under which the system experiences the average
fitness landscape, and under ``slow'' we subsume everything else,
assuming that a parameter exists that allows a smooth transition from the
``fast'' regime to the ``slow'' regime. Although
Fig.~\ref{fig:border-regions-stoch} contains
considerably less information than Fig.~\ref{fig:border-regions}, the
implications for actual landscapes are more or less the same. Most real
landscapes will have a regime that can be associated with slow
changes, and hence, we will typically observe phase diagrams of the
type of either Fig.~\ref{fig:schematic-phases12}a) or b).

As an example, consider the work of Nilsson and
Snoad~\cite{NilssonSnoad2000}, and its subsequent extension by
Ronnewinkel \etal~\cite{RonnewinkelWilkeMartinetz2000}. Nilsson and Snoad
studied a landscape in which a single peak performs a random walk through
sequence space. The peak jumps to a random neighboring position of hamming
distance 1 whenever a time interval of length $\tau_{\rm jump}$ has
elapsed. Ronnewinkel~\etal\ studied a very similar fitness landscape, but
focusing on deterministic movements of the peak that allow for the formal
definition of a quasispecies, much as for the case of periodic fitness
landscapes in Section~\ref{sec:diff-equ-formalism}. Ronnewinkel~\etal\ could
verify the results of Nilsson and Snoad on more fundamental theoretical
grounds.

The parameter $\tau_{\rm jump}$ in the jumping-peak landscape determines
whether the changes happen on a short or on a long time scale. If $\tau_{\rm
  jump}$ is very large, the landscape is static most of the time, and the
population has enough time to settle into equilibrium before the peak jumps to
a new position. If $\tau_{\rm jump}$ is very small, on the other hand, the
peak has moved away long before the population has had the time to form a
stable quasispecies. It was found that, for very fast changes, the population
fails to track the peak, and
selection breaks down. Nilsson and Snoad conclude therefore
that ``dynamic landscapes have strong constraints on
evolvability''. However, this conclusion may have to be reevaluated if
we reconsider their landscape from the viewpoint of the general theory
developed here. In a fast changing landscape, it is the
time-averaged fitness landscape which matters. In the particular example of a
randomly moving peak, this average is in general not very meaningful. However,
it allows us to render the observations of Nilsson and Snoad plausible, and to
understand under what conditions these observations would change. If we
consider an interval of length $\tau$ for which
Eq.~(\ref{eq:stoch-formal-sol}) holds, and assume that the changes in the
landscape are much faster than $\tau$, i.e., $\tau\ll\tau_{\rm jump}$, then
the matrix $\bar{\mat W}_1(t_0, \tau)$ will contain a number of peaks with
roughly the same (small) selective
advantage over the rest of the landscape. In the succeeding interval, the
matrix  $\bar{\mat W}_1(t_0+\tau, \tau)$ will have a similar structure, but
the peaks will be at different positions. In the long run, every single peak
position has on average a vanishing selective advantage over every other peak
position, and a quasispecies cannot form. However, in the rare case that the
peak by chance remains in a small region of the genotype space for a prolonged
amount of time, a momentary quasispecies will form there. Thus, in
the situation Nilsson and Snoad have studied, selection does not break down
due to the mere fact that the landscape is changing fast, but it breaks down
because the typical landscape's average over some time interval yields a highly
neutral landscape.
If the peak's movements were such that back jumps would be much more likely, or
that the peak would be confined to
a small portion of the sequence space, we would clearly see selection. This
suggests the viewpoint that the time-averaged
landscape determines the ``regions of robustness'' in the
landscape as the regions in which even fast changes in the landscape do
not destroy the quasispecies.

In addition to the complete breakdown of adaptation for a fast changing
landscape, Nilsson and Snoad observe a sharp decrease of order in their model
for very low mutation rates. On first glance, one might be tempted to explain
this observation with an asymptotically flat landscape for $R\rightarrow 0$,
as described at the end of Sec.~\ref{sec:schem-phase-diag}. However, this is
the wrong explanation in the case of a randomly moving peak. The breakdown of
adaptation for small $R$ can be observed for such large $\tau_{\rm jump}$ that
for any non-zero $R$, the landscape cannot be considered flat. The true origin
of that effect is the population's convergence onto a single point
in the genome space for these low mutation rates. Therefore, the moment the
peak position changes, there is no variance in the population that would enable
it to move over to the new peak. Note that the nature of this effect is very
different from the normal error catastrophe. The error catastrophe occurs
because replication is too erroneous to allow for information to be stored in
the sequences. As a result, a uniform population forms. In the case of small
$R$ in the moving peak landscape, however, the catastrophe occurs because the
replication is too faithful. This means in particular that no uniform
population forms. Rather, the population always grows to some extent on the
new peak position, yet
the peak does not rest long enough on that position to allow the sequences to
grow to a macroscopic concentration. For the above reasons, we propose to
refer to this effect as the convergence catastrophe, as opposed to the error
catastrophe, in order to point out that the breakdown of adaptation in that
case is not caused by erroneously copied sequences.

Another example of a stochastic fitness landscape is a noisy fitness
landscape, i.e., one in which the fitness peaks have a noise term added to
them (which means effectively that the population cannot measure the fitness
exactly), or one in which a fitness peak
has a random position. The first case has been studied by Levitan and
Kauffman~\cite{LevitanKauffman95,Levitan97} in the framework of $NK$
landscapes~\cite{KauffmanLevin87,Kauffman92}, while the second case has been
studied in the framework of population genetics by
Gillespie~\cite{Gillespie91}. In both cases, it has been observed that the
population adapts to the mean fitness, which fits nicely into the general
concepts we have presented here.

\section{Finite Populations}
\label{sec:finite-populations}

In the previous sections, we exclusively studied infinite
populations. However, as the genotype space generated
by even moderately long sequences is exponentially large (there are already
$10^{30}$ different sequences of length 100, for example) , it will be
essentially empty for any
realistic finite population. When most of the possible sequences are
not present in the population, the concentration variables become
useless, and the outcome of the differential equation formalism may be
completely different from the actual behavior of the population. For
static fitness landscapes, the effects of a finite population size are
reasonably well understood. If the fitness landscape is very simple (a
single peak landscape), the population is reasonably well described
by finite stochastic sampling from the infinite population
concentrations. Moreover, the error threshold generally moves towards
smaller $R$ with decreasing population size~\cite{NowakSchuster89}. In
a multi-peak landscape, the finite population localizes relatively
quickly around one peak, and remains there, with a dynamics similar to
that in a single-peak landscape. In the rare case that a mutant
discovers a higher peak, the population moves over to that peak,
where it remains again.
The main difference between a finite and an infinite population on a
landscape with many peaks is given by the fact that the infinite
population will always build a quasispecies around the highest peak,
whereas the finite population may get stuck on a suboptimal peak.
Above the error threshold, a finite population
starts to drift through genotype space, irrespective of the landscape.

A finite population on a dynamic landscape will of course show a
similar behavior, but in addition to that other effects come into
play that are tightly connected to the dynamics of the landscape. The
most important difference between static and dynamic landscapes is the
possible existence of a temporarily ordered phase in the latter case,
which is where we should expect the major new dynamic effects.

In the infinite population limit, the temporarily ordered phase
generates an alternating pattern of a fully developed quasispecies and
a homogeneous sequence distribution. What changes if a finite
population evolves in that phase? At those points in time when a
quasispecies
is developed, the finite population's sequence concentrations are
given by stochastic sampling from the infinite population result,
similarly to static landscapes. As soon as the quasispecies breaks
down (and this may happen earlier than the infinite population
equations predict, because the error threshold shifts to a lower
error rate for a finite population), the population starts to disperse
over the landscape. Because of that, the population may loose track of
the peak it was centered at previously. Therefore, when the system
again enters a time interval in which order should appear, the population
may not be able to form a quasispecies, thus effectively staying in
the disordered regime, or it may form a quasispecies at a different
peak. In this manner, the temporarily ordered phase can create a third
possibility for a population to leave a local peak, in addition to the
escape via neutral paths or to entropy-barrier crossing, which are
present exclusively in static landscapes~\cite{NimwegenCrutchfield99}.

\subsection{Numerical results}
\label{sec:numerical-results}

The numerical results presented below have been obtained with a
genetic algorithm on $N$ sequences per generation. We have used the
following mutation and selection scheme in order to remain as close as
possible to the Eigen model:
\begin{enumerate}
\item To all sequences $i$ in time step $t$, we assign a probability
  to be selected and mutated,
  \begin{equation}
    p_{i,\rm mut}(t)=\frac{A_i(t)}{\sum_i [1/\Delta t
      + A_i(t)-D_i(t)]n_i(t)}\,,
  \end{equation}
  and a probability to be selected but not mutated,
  \begin{equation}\label{eq:p-i-select}
    p_{i,\rm sel}(t)=\frac{1/\Delta t-D_i(t)}{\sum_i [1/\Delta t
      + A_i(t)-D_i(t)]n_i(t)}\,.
  \end{equation}
  Here, $\Delta t$ is the length of one time step, and $n_i(t)$ is the
  number of sequences of type $i$.
\item We choose $N$ sequences at random according to the above defined
  probabilities $p_{i,\rm mut}(t)$ and $p_{i,\rm sel}(t)\}$. That means, we
  perform $N$ independent drawings, and in each drawing, a sequence $i$ has a
  probability $p_{i,\rm sel}(t)\}$ to be copied faithfully into the next
  generation, and a probability $p_{i,\rm mut}(t)$ to be copied
  erroneously. The $N$ sequences such generated form the population at time
  step $t+\Delta t$.
\end{enumerate}
Note that we assume generally
\begin{equation}
  D_i(t)<\frac{1}{\Delta t}\quad\mbox{for all $i$, $t$,}
\end{equation}
so that $p_{i,\rm sel}(t)$ defined in Eq.~(\ref{eq:p-i-select}) is
always positive.

For an infinite population, such a genetic algorithm
evolves according to the equation
\begin{equation}\label{eq:basic-genal-equation}
  \vec x(t+\Delta t) = G\big(\vec x(t), t\big)\,,
\end{equation}
where $\vec x(t)$ is the vector of concentrations at time $t$, and
$G(\vec x, t)$ is the operator that maps a population at time $t$ onto
a population at time $t+1$,
\begin{equation}
  G(\vec x, t)=\frac{[\Delta t\mat W(t)+\Idmat]\vec x}
        {\vec e^\transp\cdot
          \big([\Delta t \mat A(t)-\Delta t\mat D(t)+\Idmat]\vec x\big)}\,.
\end{equation}
In Eq.~(\ref{eq:basic-genal-equation}), we can replace the non-linear operator
$G(\vec x,t)$ with a linear operator $\tilde G(\vec y, t)$,
\begin{equation}\label{eq:lin-genal-operator}
  \tilde G(\vec y, t) = [\Delta t\mat W(t)+\Idmat] \vec y\,.
\end{equation}
The linear operator acts on variables $\vec y$ that are related to the
concentrations $\vec x$ via
\begin{equation}
  \vec x(t)=\frac {\vec y(t)}{\vec e^\transp \cdot \vec y(t)}\,.
\end{equation}
By comparing Eq.~(\ref{eq:lin-genal-operator}) with
Eq.~(\ref{eq:basic-discrete}), we observe that there exists a direct
correspondence between the genetic algorithm for an
infinite population and the discrete quasispecies model. This implies in
particular that for periodic landscapes in the genetic algorithm, the
expression for the monodromy matrix $\mat X(t_0)$, Eq.~(\ref{eq:discreteX}),
is exact.

For a finite population, the operator $G(\vec x, t)$ still
determines the dynamics. However, the deterministic description
Eq.~(\ref{eq:basic-genal-equation})  has to be replaced by a
probabilistic one, namely Wright-Fisher or multinomial sampling. If
$G_i(\vec x, t)$ denotes the $i$th component of the concentration
vector in the next time step, the probability that a population $\vec
x_1=(m_1, m_2, \dots)/N, \sum_i m_i=N$, produces a
population $\vec x_2=(n_1, n_2, \dots)/N, \sum_i n_i=N$, in
the next time step, is
given by
\begin{equation}\label{eq:multinomial-sampling}
  P(\vec x_1 \rightarrow \vec x_2, t) = N!\prod_i
     \frac{G_i(\vec x_1, t)^{n_i}}{n_i!}\,.
\end{equation}
A proof that the stochastic process described by
Eq.~(\ref{eq:multinomial-sampling}) does indeed converge to the
deterministic process Eq.~(\ref{eq:basic-genal-equation}) in the limit
$N\rightarrow \infty$ has been given by van Nimwegen \etal~\cite{Nimwegenetal97a}.

\subsubsection{Loss of the master sequence}

\begin{figure}[tb]
\centerline{
 \epsfig{file={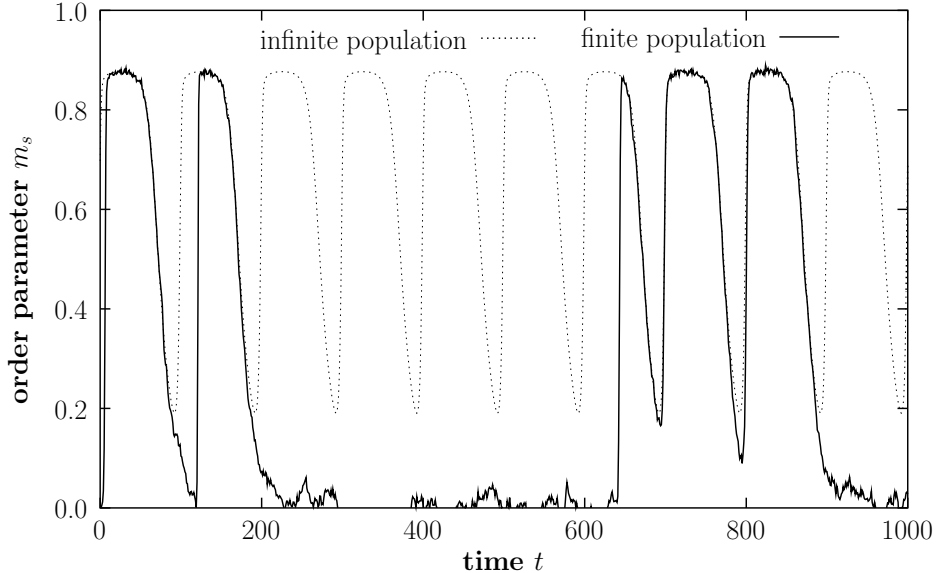}, width=0.9\columnwidth}
}
\ \\[-2ex]
\caption{\label{fig:finite-pop-l15R0.06}A single run of a population
 of $N=1000$ sequences of length $l=15$ in the oscillating Swetina-Schuster
 landscape. Parameters were
 $A_0(t)=e^{2.4}\exp(2\sin\omega t)$,  $A_i=1$ for $i>0$, $R=0.06$,  $T=100$,
 $\Delta t=1$. The dashed line indicates the theoretical result for an
 infinite population.
}
\end{figure}

Our first example of a finite population in a dynamic fitness
landscape demonstrates what happens if in the temporarily ordered
phase the master sequence is lost due to sampling fluctuations.
Fig.~\ref{fig:finite-pop-l15R0.06} depicts a run of a
finite population consisting of $N=1000$ sequences of length $l=15$,
initialized randomly at $t=0$, in an oscillating Swetina-Schuster
landscape. For comparison, we have plotted the theoretical
result for an infinite population. The infinite population is always
in an ordered state and the order parameter $m_s$ never takes on values
smaller than 0.2. Nevertheless, the finite population is likely to loose the
master sequence whenever the order parameter of the infinite
population reaches its minimum, since the error threshold is shifted
towards lower error rates for finite populations. In our example run,
the master sequence was lost at the end of the first oscillation
period, but it was rediscovered shortly afterwards, so that the
population could follow the infinite population dynamics for most of
the second oscillation period as well. Right after a loss of the
master sequence, the probability to rediscover the master has its highest
value, because the population is still centered around the master
sequence. Once the population has had the time to drift away from the
position of the master sequence, the probability of a rediscovery
drops rapidly. This is what happened at the end of the second
oscillation period. The population completely lost track of the master
sequence, and it took the population more than 4 oscillation periods
to rediscover
it. This is the main difference between a finite and an infinite
population in the temporarily ordered phase. For an infinite
population, the interval of disorder has the same well-defined length
in each oscillation period, whereas for a finite population, once the
population has entered the disordered state, it may take a long time until an
ordered state is reached again. In fact, for the case of a single peak
in a very large sequence space and a small population, the peak may
effectively be lost forever once it has disappeared from the
population.
This can be seen as a dynamic version of Muller's
ratchet~\cite{Muller64}. A trait whose advantageous influence
on the overall fitness of an individual is reduced at some point (it
is not necessary that the trait becomes completely neutral or even
deleterious) may be lost from the population due to sampling
fluctuations. If at a later stage this trait again becomes very
advantageous, it is unavailable to the population until it
is rediscovered independently. However, in most cases a rediscovery is very
unlikely.

\subsubsection{Persistency}

\begin{figure}[tb]
\centerline{
 \epsfig{file={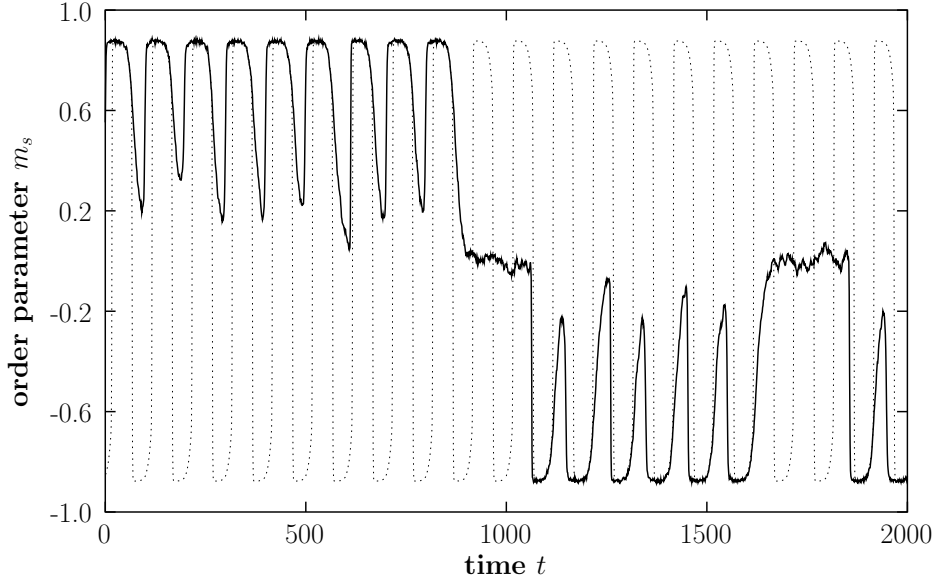}, width=0.9\columnwidth}
}
\ \\[-2ex]
\caption{\label{fig:finite-pop-2master-l15R0.06}A single run of a
 population of $N=1000$ sequences in a landscape as given in
 Eq.~(\ref{eq:two-peaks-fitness}). All parameters were identical to
 the setup of Fig.~\ref{fig:finite-pop-l15R0.06}. The dashed line
 again indicates the theoretical result for an infinite population.
}
\end{figure}

Another aspect of a finite population in a dynamic landscape is
persistency, the tendency of a finite population not to be able to follow
the changes in the landscape, even though the infinite population limit
predicts this. An example of that effect is shown in
Fig.~\ref{fig:finite-pop-2master-l15R0.06}. In that example, we have two
alternating peaks at opposite corners of the boolean hypercube, as
given by Eq.~(\ref{eq:two-peaks-fitness}). Note that the peaks'
minimal height is relatively small, but still larger than the rest of
the landscape's height. In fact, all parameters are identical to the
situation shown in Fig.~\ref{fig:finite-pop-l15R0.06}, so that this
figure can be seen as an example of the dynamics around one of the peaks in
Fig.~\ref{fig:finite-pop-2master-l15R0.06}. The infinite population
result in Fig.~\ref{fig:finite-pop-2master-l15R0.06} predicts that the
population should move to the other peak whenever it becomes
the higher one. However, the finite population does not follow this
scheme. It stays localized around one of the two peaks for a long
time, because a finite population does not
occupy all possible points in the sequence space at the same
time. Therefore, if a peak grows at a mutational distance far from the
currently occupied peak, no sequence in the population is there to
exploit the advantage, and hence the new opportunity goes
undetected. Only if the
population looses track of the first peak, which is possible because
of the temporarily ordered phase, it can discover the second peak
during its random drift. In the run of
Fig.~\ref{fig:finite-pop-2master-l15R0.06}, this has happened twice. The first
time, the population discovered the alternative
peak at the end of the drift, and the second time, it
rediscovered this same peak.

Let us compare the case of a finite population in a dynamic landscape with
several growing and shrinking peaks to a rugged, but static landscape. In the
latter case, once the population
has reached a local optimum it remains there, unless a rare mutation
opens the possibility to move to a new, higher peak. The same applies
to the dynamic situation. But in addition, the fluctuations and
oscillations of the fitness values destabilize the population on local
optima, and allow it to continue its search for other local optima. If
the landscape's dynamics is such that the population, by following the
local optima, moves into regions of low average fitness (observed,
e.g., in~\cite{Wilke99}), the landscape might be called ``deceptive'',
and in the opposite case, it might be called ``well-behaved''.

\subsection{A finite population on a simple periodic fitness landscape}
\label{sec:ana-finite-pop}

\subsubsection{The error tail approximation}

In the above examples, we saw that the time it takes until the master sequence
is rediscovered after it has been lost in the temporarily ordered phase
may be much larger than the period length of the landscape. Hence, for
several periods, the population does not follow the infinite population
results, but remains in a disordered state. It would be desirable to
have an analytic description of this behavior, and, in particular, to
have an estimate of the probability with which a complete period is
skipped, i.e., with which the master sequence is missed for a whole
oscillation period. Unfortunately, the continuous time dependency of
the master sequence's replication rate used in Sec.~\ref{sec:numerical-results},
\begin{equation}\label{eq:againA0oft}
  A_0(t)=A_{0, \rm stat}\exp(\epsilon \sin\omega t)\,,
\end{equation}
renders the corresponding
calculations very complicated. Instead, we study in this
section a simplified fitness landscape that displays a temporarily
ordered phase similar to Fig.~\ref{fig:finite-pop-l15R0.06}, but
that is much easier to handle analytically. For a fitness landscape
such as Eq.~(\ref{eq:againA0oft}), we can---for sufficiently high
error rate $R$---divide the oscillation period into two
intervals. During the first interval $I_1$, of length $T_1$, the
population is in an ordered state provided that the master sequence is
present in the population, and during the second interval $I_2$, of
length $T_2$, the population is in a disordered state, even if the
master sequence is present. The beginning of the first interval need
not coincide with the beginning of the oscillation period, but after a
suitable shift of the time origin, this is always the case. Note that
for a finite population, the second interval is larger than predicted
by the infinite population limit, and it may exist even if the
infinite population limit predicts a length $T_2=0$, because the error
threshold is shifted towards smaller error rates for finite
populations~\cite{NowakSchuster89,Wieheetal95}, as discussed previously.

Our approximation here is to keep the fitness landscape constant
during the intervals $I_1$ and $I_2$. During the interval $I_1$, we
let the
master sequence replicate with rate $A_0\gg1$, while all other sequences
replicate with $A=1$. During the second interval on the contrary,
the fitness landscape becomes flat, i.e., all sequences replicate with
$A=1$. We continue to study the discrete process and set $\Delta t=1$,
so that $T_1$ and $T_2$ give the number of time steps spent in each
interval. In summary, the replication rate $A_0(t)$ satisfies
\begin{equation}
  A_0(t) = \left\{ \begin{array}{c@{:\qquad}l}
      a & \phi \leq T_1 \\
      1 & \mbox{else}\,.
    \end{array}\right.
\end{equation}

In order to obtain expressions that can be treated easily even for a
finite population, we use the error tail approximation introduced
in~\cite{NowakSchuster89}. In that approximation, the state of the
system is fully described by the concentration of the master
sequence. All other sequences are assumed to be uniformly spread over
the remaining genotype space. This approximation underestimates the
mutational backflow into the master sequence, and hence it
underestimates the concentration of the master sequence itself, but this small
deviation can be accepted in the light of the enormous simplifications
in the calculations.

Before studying the finite population dynamics, let us turn
quickly to the infinite population limit. We express the state of
the system at time $t$ by a vector $\vec x(t)=(x_0(t), x_1(t))^\transp$,
where $x_0(t)$ is the concentration of the master sequence, while
$x_1(t)=1-x_0(t)$ stands for the total concentration of all other
sequences. The generation operator $G(\vec x, t)$ maps the population
at time $t$ into the population at time $t+1$, i.e.,
\begin{equation}
  \vec x(t+1) = G\big(\vec x(t), t\big)\,.
\end{equation}
Here, $G(\vec x, t)$ is given by
\begin{equation}\label{eq:def-operator-G}
  G(\vec x, t) = \frac{[\mat Q\mat A(t) + \Idmat]\vec x}{A_0(t)x_0+x_1+1}\,.
\end{equation}
$\mat Q$ is the $2\times 2$ matrix
\begin{equation}
  \mat Q = \left( \begin{array}{cc}
      (1-R)^l & \frac{1-(1-R)^l}{2^l-1} \\
      1-(1-R)^l & 1-\frac{1-(1-R)^l}{2^l-1}
      \end{array} \right)\,,
\end{equation}
and $\mat A(t)=\diag (A_0(t), 1)$. The linear operator
$\tilde{\mat G}(t) = \mat Q\mat A(t)+\Idmat$ describes the evolution
of the variables $\vec y(t)$,
\begin{equation}
  \vec y(t+1) = \tilde{\mat G}(t)\vec y(t)\,,
\end{equation}
which map into the original variables via
\begin{equation}
  \vec x(t) = \frac{\vec y(t)}{\vec e^\transp\cdot \vec y(t)}\,,\qquad \vec e^\transp
  = (1, 1)\,.
\end{equation}
Hence, the eigensystem of $\tilde{\mat G}$ fully describes the time
evolution of $\vec x(t)$. For the eigenvalues of $\tilde{\mat G}$, we
find
\begin{equation}\label{eq:tildeGeigenvalues}
  \lambda_{0,1} = \frac{1}{2}\left[\tilde G_{00}+\tilde G_{11} \pm
        \sqrt {(\tilde G_{00}-\tilde G_{11})^2 +
          4\tilde G_{01}\tilde G_{10}}\right]\,,
\end{equation}
where the plus sign corresponds to the index 0, and the minus sign
corresponds to the index 1. The eigenvectors are
\begin{align}\label{eq:tildeGeigenvectors}
  \vec \phi_{0,1}&= \frac {1}{1+\xi_{\pm}}\left (1, \xi_{\pm}\right)^\transp\,,\\
\intertext{with}
\label{eq:tildeGxi}
\xi_{\pm} &= \frac{\tilde G_{00}-\tilde G_{11}}{2\tilde G_{01}} \pm
       \frac{1}{\tilde G_{01}}
       \sqrt {\frac{1}{4}(\tilde G_{00}-\tilde G_{11})^2 +
          \tilde G_{01}\tilde G_{10}}\,.
\end{align}
Of course, the eigenvalues and the eigenvectors are different for the
two intervals $I_1$ and $I_2$. For the first interval, inserting the
explicit expressions for $\tilde G_{ij}$ into
Eqs.~(\ref{eq:tildeGeigenvalues})--(\ref{eq:tildeGxi}) does not lead
to a substantial simplification of the expressions. For the second interval,
however, we find for the eigenvalues
\begin{subequations}
  \begin{align}
    \lambda^{(2)}_0 &= 2\,,\\
    \lambda^{(2)}_1 &= 2-\frac{1-(1-R)^l}{1-2^{-l}}\,,
  \end{align}
\end{subequations}
and for the eigenvectors
\begin{subequations}
  \begin{align}
    \vec\phi^{(2)}_0 &= (2^{-l}, 1-2^{-l})^\transp\,,\\
    \vec\phi^{(2)}_1 &= (1, -1)^\transp\,.
  \end{align}
\end{subequations}
The superscript $(2)$ indicates that these results are only valid for
the interval $I_2$.
From the above expressions, we obtain a simple formula for the evolution
of the master sequence's concentration during the interval $I_2$ in the
following manner. Let the
interval start at time $t$, and let the concentration of the master sequence at
that moment in time be $x_0(t)$. Then we find $n$ time steps later
\begin{equation}\label{eq:sol-tildeG-errortail}
  x_0(t+n) = \frac{\alpha_0 \vec\phi^{(2)}_0 + \alpha_1
    \left(\lambda^{(2)}_1/\lambda^{(2)}_0\right)^n \vec\phi^{(2)}_1}
  {\alpha_0 (\vec e^\transp\cdot \vec\phi^{(2)}_0) + \alpha_1
   \left(\lambda^{(2)}_1/\lambda^{(2)}_0\right)^n (\vec e^\transp\cdot \vec\phi^{(2)}_1)}\,,
\end{equation}
where $\alpha_0$ and  $\alpha_1$ have to be chosen such that
\begin{equation}\label{eq:cond-on-alpha}
  x_0(t) = \alpha_0 \vec\phi^{(2)}_0 + \alpha_1 \vec\phi^{(2)}_1\,.
\end{equation}
After solving Eq.~(\ref{eq:cond-on-alpha}) for $\alpha_0$ and
$\alpha_1$ and inserting everything back into
Eq.~(\ref{eq:sol-tildeG-errortail}), we find
\begin{equation}\label{eq:x0-decay-flat-landscp}
  x_0(t+n) = 2^{-l} +\left[x_0(t)-2^{-l}\right]\left(1 -
    \frac{1-(1-R)^l}{2(1-2^{-l})}\right)^n\,.
\end{equation}
This formula is sufficiently close to the solution obtained from
diagonalization of the full $2^l\times 2^l$ matrix $\mat Q$ in a flat
landscape, and can be considered a good approximation to the actual
infinite population dynamics~\cite{Ronnewinkel99}. In
principle, a similar formula can be derived for the interval $I_1$,
but again, the expressions become very complicated and do not lead to
any new insight.

Equation (\ref{eq:x0-decay-flat-landscp}) demonstrates that a
macroscopic proportion of the master sequence that may have built up
during the interval $I_1$ quickly decays to the expected concentration
in a flat landscape, $2^{-l}$.

Let us now study finite population corrections. We assume the duration of the
interval $I_1$ is long enough so that the quasispecies can
form. The asymptotic concentration of the master sequence can then be
calculated from a birth and death process as done
in~\cite{NowakSchuster89}. The alternative diffusion approximation
used in~\cite{Wieheetal95} is of no use here because it allows only
replication rates $A_0$ of the form $A_0=1+\epsilon$ with a small
$\epsilon$~\cite{Ewens79}. In~\cite{NowakSchuster89}, the probabilities
$p_k$ to find the master sequence $k$ times in the asymptotic
distribution are given by
\begin{equation}
  p_k=\frac{\tilde p_k}{\sum_{i=0}^N\tilde p_i}\quad \mbox{with
  $\tilde p_k=\frac{\mu^+_{k-1}}{\mu^-_k}\tilde p_{k-1}$ and $\tilde p_0=1$.}
\end{equation}
The probabilities $\mu^+_i$ and $\mu^-_i$ read here
\begin{equation}
  \mu^+_i = \frac{N-i}{N}
  \left(\left[\tilde G^{(1)}_{00}-1\right]\frac{i}{N}
      + \tilde G^{(1)}_{01}\frac{N-i}{N}\right)
\end{equation}
and
\begin{equation}
  \mu^-_i = \frac{i}{N}
  \left(\tilde G^{(1)}_{10}\frac{i}{N}
      + \left[\tilde G^{(1)}_{11}-1\right]\frac{N-i}{N}\right)\,.
\end{equation}
The expected asymptotic concentration becomes
\begin{equation}\label{eq:expected-asymp-conc}
  x_0(\infty) = \frac{1}{N}\sum_{k=0}^{N}k\,p_k\,.
\end{equation}
Unfortunately, there exists no analytic expression for
$x_0(\infty)$. However, its value is easily computed numerically. With the
above assumption about the length of the interval $I_1$, we can
suppose that at the end of $I_1$ the concentration of $x_0$ is given
by $x_0(\infty)$. During the interval $I_2$, the concentration of the
master sequence will then decay.

\subsubsection{The probability to skip one period}

If at the end of the interval $I_2$ the master sequence has been lost
because of sampling fluctuations, and if in addition to that the
correlations in the population have decayed so far that we can assume
maximum entropy, what is the probability that the master sequence is
rediscovered in the following interval $I_1$? The process of
rediscovering the master consists of two steps. The master sequence
has to be generated through mutation, and then it has to be fixated in
the population, i.e., it must not be lost again due to sampling
fluctuations. First, we calculate the probability
$P_{\rm miss}$ that the master is not generated in one time step. This
corresponds to the probability that the multinomial sampling of the
operator $G^{(1)}(\vec x)$ maps a population $\vec x=(0,1)^\transp$ into
itself. Hence, we have
\begin{align}\label{eq:P-miss-expr}
  P_{\rm miss}&=N!\prod_{i=0}^1 \frac{G_i^{(1)}(\vec x)^{n_i}}{n_i!}\notag\\
   &= \left(\frac{Q_{11}+1}{2}\right)^N=
   \left[1-\frac{1-(1-R)^l}{2^{l+1}-2}\right]^N\,.
\end{align}
$G_i^{(1)}(\vec x)$ stands for the $i$th component of the outcome of
$G^{(1)}(\vec x)$.

The probability that the master sequence becomes fixated requires more
work. Denote by $\pi(x, t)$ the probability that the master
sequence has reached its asymptotic concentration at time $t$, given
that it had the initial concentration $x$ at time $t=0$. The
asymptotic concentration is given by $x_0(\infty)$ defined in
Eq.~(\ref{eq:expected-asymp-conc}). Then, the probability $\pi(x,t)$
satisfies to second order the backward Fokker-Planck equation
\begin{equation}\label{eq:backwardFokkerPlanck}
  \frac{\partial\pi(x,t)}{\partial t} = \langle dx_0\rangle
  \frac{\partial\pi(x,t)}{\partial x}
  + \frac{\langle (dx_0)^2\rangle}{2}
  \frac{\partial^2\pi(x,t)}{\partial x^2} \,.
\end{equation}
The moments $\langle dx_0\rangle$ and $\langle (dx_0)^2\rangle$ can be
calculated along the lines of~\cite{Nimwegenetal97a},
and we find
\begin{align}
  \langle dx_0\rangle&= \left(\frac{1}{2}\lambda_0^{(1)}-1\right) x_0
   =: \gamma x_0\,,\\
  \langle (dx_0)^2\rangle& = \frac{x_0(1-x_0)}{N}\,.
\end{align}
The solution to Eq.~(\ref{eq:backwardFokkerPlanck}) for
$t\rightarrow\infty$ is then obtained as in~\cite{Nimwegenetal97a},
and we find
\begin{align}\label{eq:pi-infty}
  \pi_\infty&:=\pi\left(\frac{1}{N}, \infty\right)
     =\frac{1-\left(1-\frac{1}{N}\right)^{2N\gamma+1}}
     {1-\left(1-x_0(\infty)\right)^{2N\gamma+1}}\\\label{eq:pi-infty-approx}
     &\approx 1-e^{-2\gamma}\,.
\end{align}
As the initial concentration of $x_0$, we have used $1/N$, since it
is---for the parameter settings we are interested in---extremely
unlikely that more than one master sequence is generated in one time
step. The approximation in the second line is only valid for large
population sizes. It generally underestimates the true value of $\pi_\infty$.

Note that the expression for  $\pi_\infty$ given in
Eq.~(\ref{eq:pi-infty}) reaches the value 1 for the (relatively large)
error rate $R$ close to the error threshold for which
$x_0(\infty)=1/N$. Naively, one would assume that $\pi_\infty$ decays
with increasing error rate, since mutations increase the risk that
good traits are lost, and indeed the approximate expression in
Eq.~(\ref{eq:pi-infty-approx}) decays with increasing error
rate. However, since $\pi_\infty$ is the probability that the master
sequence reaches its equilibrium concentration, and the equilibrium
concentration vanishes close to the error threshold, $\pi_\infty$ must
rise to 1 at the error threshold.

\begin{figure}[tb]
\centerline{
 \epsfig{file={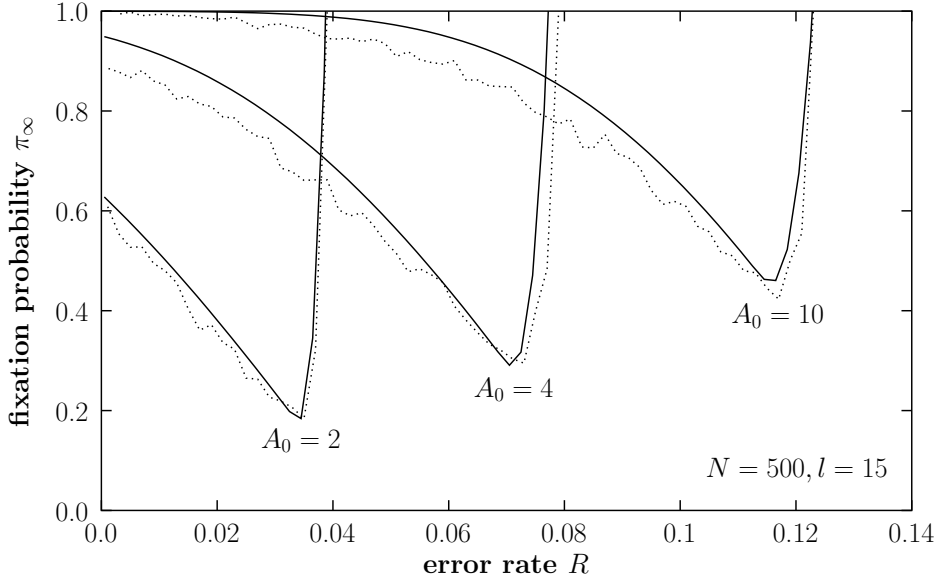}, width=0.9\columnwidth}
}
\ \\[-2ex]
\caption{\label{fig:pi}The fixation probability $\pi_\infty$ as a
 function of the error rate $R$ for three different heights of the
 peak. The solid lines stem from the analytic expression
 Eq.~(\ref{eq:pi-infty}),
 and the dotted lines stem from measurements on a finite population
 consisting of $N=500$ sequences.
}
\end{figure}

We have performed simulations with a finite population to test the
validity of Eq.~(\ref{eq:pi-infty}). For a number of runs, we have
initialized the population at random, but with exactly one instance of
the master sequence, and have counted how often the master's
concentration reached $x_0(\infty)$ and how often it reached 0. The
result of these runs are shown in Fig.~\ref{fig:pi}. Clearly,
numerical and analytical results are in good agreement.

Finally, we need an estimate of the time $\tau$ it takes from the time
the master sequence is discovered to the time in which the
equilibrium concentration is reached for the first time. We
again follow the calculations in~\cite{Nimwegenetal97a}, and assume that the
process of fixation can be treated in the infinite population
limit. From Eq.~(\ref{eq:def-operator-G}), we obtain for the change in
the variable $x_0(t)$ during one time step in the interval $I_1$
\begin{equation}
  x_0(t+1)-x_0(t) = \frac{-(a-1)x_0(t)^2+(Q_{00} a - Q_{01} - 1)x_0(t)
       + Q_{01}}{(a-1)x_0(t)+2}\,.
\end{equation}
This can be approximate with a differential equation,
\begin{equation}
  \frac{dx_0(t)}{dt} \approx x_0(t+1)-x_0(t)\,,
\end{equation}
which we can solve for $t$ as a function of $x_0$ to obtain
\begin{align}
  t&= \frac{b+4}{z}\left({\rm Atanh}\frac{b-2sx_0}{z}-{\rm Atanh}\frac{b-2s/N}{z}\right)\notag\\
 &\qquad\qquad\qquad\qquad\qquad-\frac{1}{2}\ln\frac{-sx_0^2 + bx_0 + Q_{01}}{-s/N^2 +
  b/N + Q_{01}}\,,
\end{align}
with
\begin{align}
  s&=a-1\,,\\ b&=Q_{00} a -Q_{01} - 1\,,
\end{align}
and
\begin{equation}
  z=\sqrt{4sQ_{01}+b^2}\,.
\end{equation}
Therefore, for the estimated time it takes until the master sequence
becomes fixated we will use in the following
\begin{equation}\label{eq:fixation-time}
  \tau = t\big(x_0(\infty)\big)\,,
\end{equation}
with $x_0(\infty)$ given in Eq.~(\ref{eq:expected-asymp-conc}).

We can now calculate the probability that the population skips
a whole period, i.e., that it does not find and fixate the master
during one interval $I_1$. The probability that the master sequence
has concentration zero at the beginning of the interval $I_1$ is
$(1-1/2^l)^N$. Therefore, the probability that the master sequence is
not fixated in the first time step is
\begin{equation}
  1-\left[1-\left(1-\frac{1}{2^l}\right)^N\right]\pi_\infty\,.
\end{equation}

The probability that the master sequence is not found and
subsequently fixated is given by
\begin{equation}
  1-(1-P_{\rm miss})\pi_{\infty}\,.
\end{equation}
Now, if the master sequence is found, it will roughly take the time
$\tau$ given in Eq.~(\ref{eq:fixation-time}) until the equilibrium
concentration is reached. Therefore, if the master sequence is not
found during the first
$T_1-\tau$ time steps, it normally will not reach the equilibrium
concentration in that period. Therefore, in order to calculate
the probability $P_{\rm skip}(T_1)$ that the whole interval $I_1$ is
skipped, we have to consider only the first $T_1-\tau$ time steps of
$I_1$. In case that $T_1<\tau$, we have $P_{\rm skip}(T_1)\approx
1$. The equality is only approximate, because $\tau$ is the average
time until fixation occurs. In rare cases, the fixation may happen
much faster.

Of the $T_1-\tau$ time steps, the first one is different because during
that time step we do not know whether the master sequence is present
or not, whereas for the remaining $T_1-\tau-1$ time steps we may
assume that the master sequence is not present if fixation has not
occurred. Therefore, we find
\begin{align}\label{eq:P-skip-ana}
  P_{\rm skip}(T_1)&=
  \left(1-\left[1-\left(1-\frac{1}{2^l}\right)^N\right]\pi_\infty\right)
  \left[1-\left(1-P_{\rm miss}\right)\pi_{\infty}\right]^{T_1-\tau-1}\\
&\approx
  \frac{1-\left[1-\left(1-\frac{1}{2^l}\right)^N\right]\pi_\infty}
  {1-\left(1-P_{\rm miss}\right)\pi_{\infty}} \exp\left[-(T_1-\tau)
  \left(1-P_{\rm miss}\right)\pi_\infty\right]\,.
\end{align}

\begin{figure}[tbp]
\begin{center}
\noindent\epsfig{file={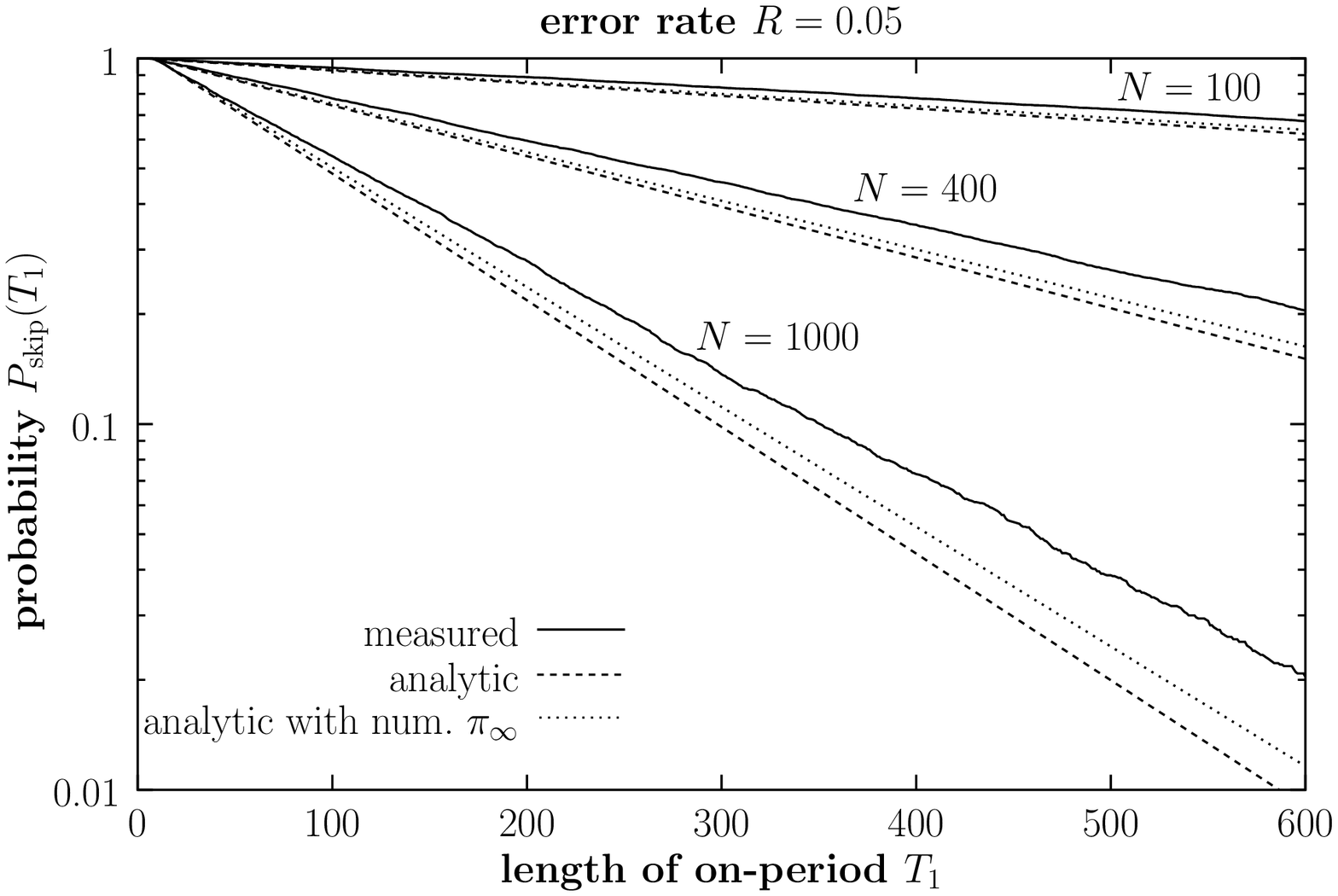}, width=0.9\columnwidth}\\
\bigskip

\noindent\epsfig{file={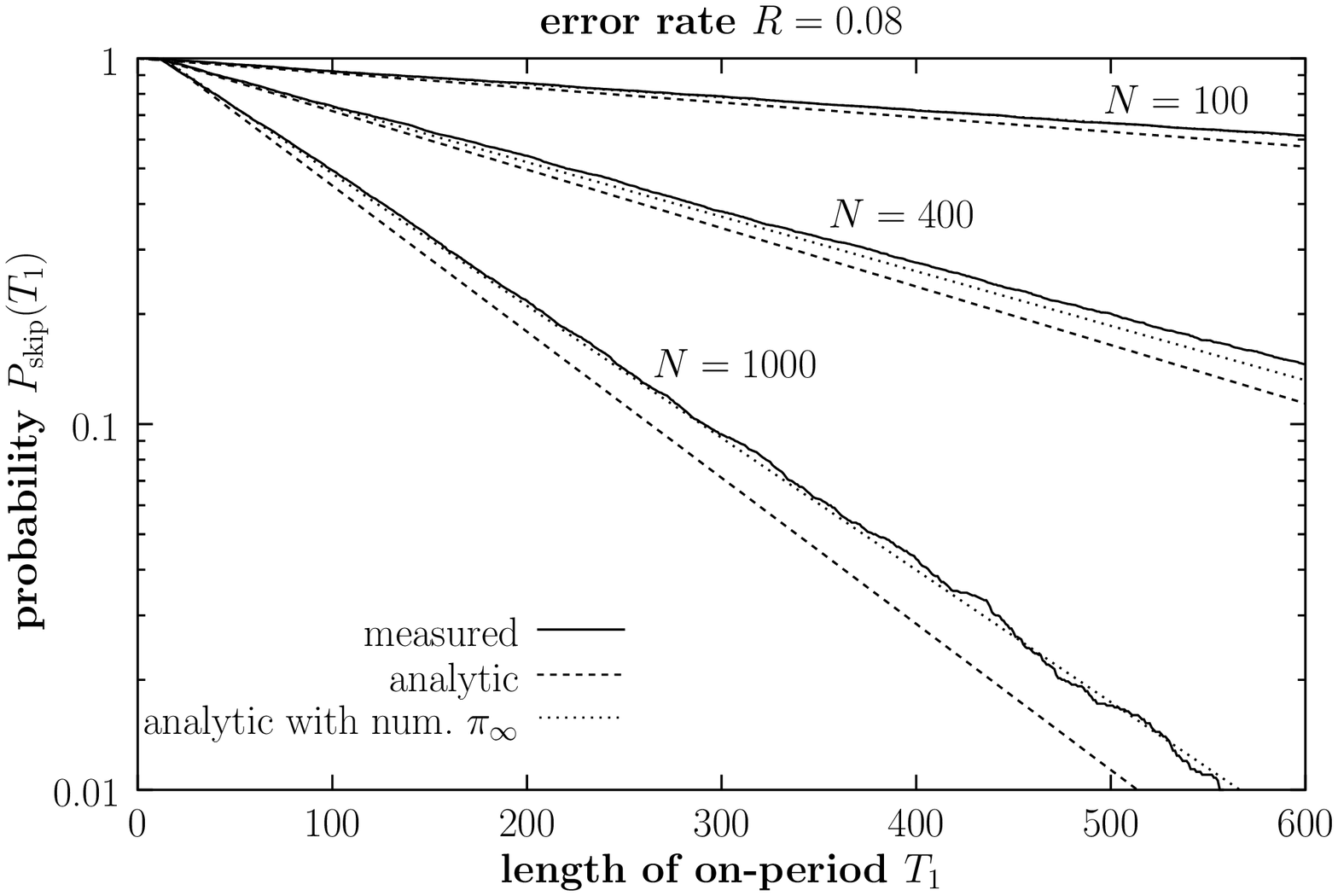}, width=0.9\columnwidth}
\end{center}
\ \\[-2ex]
\caption{\label{fig:p-find-master-l15}The probability
 $P_{\rm skip}(T_1)$ that the population skips a whole period without
 fixating the master sequence, as a function of the length of the
 interval $I_1$, for several different settings of $N$ and $R$. The
 string length is $l=15$.
}
\end{figure}

Figure~\ref{fig:p-find-master-l15} shows a comparison between this
result and numerical simulations. The simulations were carried out by
letting a randomly initialized population evolve in a flat landscape
for 100 generations, and then recording the time it took the
population to find and fixate a peak that was switched on in
generation 101. We observe that the analytic expression for
$P_{\rm skip}(T_1)$ predicts the right order of magnitude and the
right functional dependency on $T_1$, but that it generally
underestimates the exact
value. Since Eq.~(\ref{eq:P-skip-ana}) contains three quantities for
which we have only approximative expressions, namely $P_{\rm miss}$,
$\pi_{\infty}$, and $\tau$, at first it is not clear from where these
discrepancies arise. However, a systematic check quickly reveals the
main cause of the discrepancies. First of all, note that $\tau$ merely
shifts the curve to the right. Since the measured and the analytic
curves reach the value 1 at very much the same positions in
Fig.~\ref{fig:p-find-master-l15}, we can assume that $\tau$, as given
by Eq.~(\ref{eq:fixation-time}), is accurate enough for our purposes
here. Now consider the quantity $\pi_{\infty}$. In Fig.~\ref{fig:pi}, we
saw that our expression for $\pi_{\infty}$ generally gives a good
estimate of the true value, but that there are some deviations. To
check whether these deviations are responsible for the discrepancies
visible in Fig.~\ref{fig:p-find-master-l15}, we show additionally
$P_{\rm skip}(T_1)$ with $\pi_{\infty}$ determined numerically. We find that
using a numerical
$\pi_{\infty}$ enhances the prediction of Eq.~(\ref{eq:P-skip-ana}), in
particular for larger error rates. For small error
rates, however, the discrepancy is still sizable. Moreover, the analytic
expression is generally getting worse for smaller error rates. We can thus
conclude that the main problems arise from the expression for
$P_{\rm miss}$, Eq.~(\ref{eq:P-miss-expr}). Indeed, we have derived
$P_{\rm  miss}$ under a maximum entropy assumption, i.e., we
assumed that all mutants are distributed homogeneously over
sequence space. Under this assumption, the probability to find the
master is exactly the same at every time step. But in reality, the
population collapses very rapidly, even in a neutral landscape, and
then moves about as a cluster whose radius is determined by the error
rate. This introduces very long-range temporal correlations in a
population evolving in a flat landscape~\cite{DerridaPeliti91}. In
particular, for small error rates the cluster is very small, and this
can increase the probability $P_{\rm  miss}$ substantially. Note that
this effect corresponds to the underestimation of epoch durations that
van Nimwegen \etal\ found in their analysis of the Royal Road genetic
algorithm~\cite{Nimwegenetal97a}. An exact treatment of this effect
would probably have to occur along the lines
of~\cite{DerridaPeliti91}. Unfortunately, we cannot simply use their
expressions here, because of the term $+\Idmat$ present in our
definition of the operator $G(\vec x, t)$
[Eq.~(\ref{eq:def-operator-G})].

In order to check the hypothesis that the violation of the maximum
entropy condition causes the main discrepancies shown in
Fig.~\ref{fig:p-find-master-l15}, we performed additional simulations
in which we dispersed the population ``by hand'' over the complete
sequence space in every time step in which the
master sequence was not discovered. With this setup, we found a very
good agreement between the numerical and the analytical results.

What can we conclude from Eq.~(\ref{eq:P-skip-ana})? First, note that the true
$P_{\rm skip}(T_1)$ must always be larger than the value predicted by
Eq.~(\ref{eq:P-skip-ana}), because the deviations from that value are caused
by the population's collapse into a small cluster. Hence,
Eq.~(\ref{eq:P-skip-ana}) is a lower bound on $P_{\rm skip}(T_1)$, and
rediscovery of the peak is less likely than what Eq.~(\ref{eq:P-skip-ana})
predicts. According to our prediction, $P_{\rm skip}(T_1)$ decays
exponentially. This means that the probability to find the peak in one
oscillation period,
\begin{equation}
  P_{\rm find}(T_1) = 1 - P_{\rm skip}(T_1),
\end{equation}
approaches 1 for large $T_1$. This is due to the fact that the peak
will certainly be rediscovered if only we wait long enough. However,
the model we are studying here is that of a peak that is switched on
and off alternatingly, and for which each ``on''-period is of fixed
length $T_1$. In that case, the
probability to rediscover the peak within one oscillation period can  be
extremely small, as we are going to see now.  $P_{\rm skip}(T_1)$
decays with a rate of
$(1-P_{\rm miss})\pi_\infty$. We can neglect $\pi_\infty$ here, as it
is of the order of one. Then, the decay rate is for fixed $N$ and
large $l$ is approximately given by
\begin{equation}
  1-P_{\rm miss} \approx N\frac{1-(1-R)^l}{2^{l+1}}\,,
\end{equation}
i.e., it decays as $2^{-l}$. This implies in turn that already for
string lengths of 50--60 (which can be considered a rough lower bound
for typical DNA sequence lengths) and moderate $N$ and $R$, we find
$P_{\rm find}(T_1)\approx 0$ for moderate $T_1$. Hence, in many cases
it is extremely unlikely that the peak is rediscovered at all.

The above conclusion is of course tightly connected to the fact that
we have studied a landscape with a single advantageous sequence. In
the other extreme of a wide-peak (Mount Fujiama) landscape, in which the
population can sense the peak from every position in the sequence
space, the conclusions would be different. Note, however, that
neither the single-sharp-peak landscape nor the wide-peak landscape are
realistic landscapes. In a realistic, high-dimensional rugged
landscape, it is probably valid to assume that local optima, once they
are lost from the population, are never rediscovered. In such
situations, dynamic fitness landscapes can induce the loss of a local
optimum, and thus, they can accelerate Muller's ratchet\cite{Muller64}
like effects.

\section{Conclusions}
\label{sec:conclusions}

In this report, we have derived several very general
results about landscapes with periodic time dependency. First of all, a
quasispecies can be defined by means of the monodromy matrix. This
means that after a sufficiently long time, the state of the system
depends only on the phase $\phi = (t\mod T)/T$ of the oscillation, but
not on the absolute time $t$. Therefore, in
periodic fitness landscapes, the quasispecies is not a fixed mixture
of sequence concentrations. Instead, it is a $T$-periodic function of
mixtures of sequence concentrations. We have given an expansion of the
monodromy matrix in terms of the oscillation period $T$, which leads
to an extremely simple description of the system for very high
oscillation frequencies. Namely---if we assume the mutation matrix
remains constant at all times---the time-averaged fitness landscape
completely determines the behavior of the system, which is essentially
indistinguishable from a system in a static landscape. This leads to the
important conclusion that selection never ceases to act, no matter
how fast the landscape changes. The only exceptions to this rule are
due to dynamic landscapes that have a completely flat
average. In that case, the system for very fast changes behaves as
being subjected to a flat landscape, which is indistinguishable from the
behavior of a system above the error threshold. Therefore, if the
average landscape is flat, selection will break down if the changes
occur with a frequency higher than some critical frequency
$\omega^*=2\pi/T^*$. For very slow changes, on the other hand, the
system is virtually in
equilibrium all the time. This leads, in general, to a time dependent
error threshold $R^*(t)$. For mutation rates $R$ such that $\min_t
R^*(t) < R < \max_t R^*(t)$, the system is below the error threshold
for some times $t$, and it is beyond the error threshold for other
times. We have dubbed this region of the parameter space the
\emph{temporarily ordered phase}, as we see alternating patterns of order and
disorder in that phase (in the infinite population limit).
We found these general considerations to be in complete agreement with
all example landscapes that we studied.

Periodic fitness landscapes can be fully understood from the knowledge of the
monodromy matrix. Therefore, in future work it should be tried to obtain an
improved understanding of the properties of that matrix. In particular, an
expansion of that matrix in the error rate $R$ would help to further develop
the schematic phase diagrams introduced in Sec.~\ref{sec:schem-phase-diag}.

While the molecular
concentrations become $T$-periodic for $t\rightarrow\infty$ in the
infinite population limit, this is not necessarily the case when we
consider finite populations. In the temporarily ordered phase, after a
population has made the transition to the disordered state, it may not
transition back to order as the infinite
population would. Rather, once the population
has lost the ordered state, it is often difficult for the population to return
to it. From a very simple analytical model, we found that
the probability that the ordered state is not rediscovered in one
oscillation period decays exponentially with the length of the interval
in which order is possible at all. The decay constant, however, is
extremely small for large $l$, and therefore the rediscovery can
become very unlikely. In more complex landscapes, this can lead to an
acceleration of Muller's ratchet.

For the case of non-periodic landscapes, we have argued that the main
conclusions remain valid, even if our mathematical formalism is not
generally applicable
in that case. Fast changes in the landscape will average out, whereas
slow changes lead to a quasistatic adaption of the quasispecies to the
current landscape.

Throughout this report, we have assumed that mutations arise in the
copy process. An equally valid assumption is that of mutations arising
on a per-unit-time basis (cosmic ray mutations), as opposed to the per
generation basis implied by copy mutations. With the latter
assumption, one has to study the parallel mutation and selection
equations~\cite{BaakeGabriel99} instead of Eigen's equations. Since
these equations can be linearized in the same way as the quasispecies
equations, the formalism we developed applies to these
equations also. The only difference between the two types of equations is
that in the case of cosmic ray mutations, the mutation matrix $\mat Q$ and the
replication matrix $\mat A$ are added, whereas in the quasispecies
case they are multiplied.

A question that must remain unanswered within the current body of knowledge is
to what extent dynamic fitness landscapes help with the progress of
evolution. In Sec.~\ref{sec:finite-populations}, we have seen that the
dynamics of a fitness landscape can destabilize a population on a local
peak. On the one hand, being trapped in a local optimum is regarded as one
of the main hindrances to the progress of evolution, so that the destabilizing
effect seems to advance evolution. On the other hand, the same effect can lead
to the loss of an advantageous trait. Whether the positive or the negative
aspect prevails depends most certainly on details of the landscape. In a
study of adaptive  walks on dynamic $NK$ landscapes, exactly this question was
addressed~\cite{WilkeMartinetz99a}. In that particular case, it was found that
for a rapidly changing landscape, the loss of traits was dominant, whereas
a slowly changing landscape could lead to a more efficient exploration of the
high-fitness regions of genotype space. Apart from this particular study,
however, the amount to which a dynamic landscape can advance the progress of
evolution is unknown, and deserves more attention in future work.

\begin{ack}
We thank Chris Adami for carefully reading this manuscript. Part of this work
was supported by the NSF under contract DEB-9981397.
\end{ack}

\begin{appendix}
\section[High-frequency expansion of $\mat X(t)$]{High-frequency
  expansion of $\mat X(t)$ for a single oscillating fitness peak}
\label{app:high-frequ-ex}
Eq.~(\ref{eq:X-Neumann-exp}), gives an expansion of the
monodromy matrix for periodic landscapes, $\mat X(t_0)$, in terms of
the period length $T$. Here, we calculate the expansion
explicitly up to second order for an example landscape. We choose a landscape
with a single oscillating peak. The replication rates are
\begin{subequations}\label{eq:simple-single-peak}
\begin{align}
  A_0(t)&= a+b\sin (\omega t)\,,\\
  A_i(t)&= 1 \qquad \mbox{for all $i>0$}\,.
\end{align}
\end{subequations}

In this particular example, we set the decay rates to zero. With vanishing
decay rates, the
matrix $\mat W(t)$ reduces to $\mat Q\mat A(t)$, and as a
consequence, we can write the $n$th average $\overline{\mat W}_k(t)$ as
\begin{equation}
  \left(\overline{\mat W}_k(t)\right)_{ij} = \sum{\nu_1}\sum{\nu_2}\cdots
           \sum{\nu_{k-1}} Q_{i\nu_1} Q_{\nu_1\nu_2}\cdots Q_{\nu_{k-1}j}
           \overline A_{\nu_1, \nu_2,\dots \nu_{k-1},j}(t)
\end{equation}
with the generalized replication coefficients
\begin{multline}
  \overline A_{\nu_1,\dots \nu_{k-1},j}(t) =  \frac{1}{T^k}\int_0^T
      A_{i\nu_1}(t_0+\tau_1) \cdots\\\cdots \int_0^{\tau_{k-2}} A_{\nu_{k-1}}(t_0+\tau_{k-1})
      \int_0^{\tau_{k-1}} A_{j}(t_0+\tau_k)
         d\tau_1\cdots d\tau_k\,.
\end{multline}
For the landscape given in Eq.~(\ref{eq:simple-single-peak}), the
first order tensor of the generalized replication coefficients has two
independent elements, which are (assuming $i>0$)
\begin{subequations}
\begin{align}\label{eq:expans-in-T-i}
  \overline A_{0}(t)&= a\,,\\
  \overline A_i(t)&= 1\,.
\end{align}\end{subequations}
The second order tensor has four independent entries. After some
algebra, we obtain (assuming again $i>0$)
\begin{subequations}\label{eq:second-order-expans}
\begin{align}\label{eq:expans-in-T-ii}
  \overline A_{00}(t) &= \frac{a^2}{2}\,,\\
  \overline A_{0i}(t) &= \frac{a}{2}-\frac{b}{2\pi}\cos(\omega t)\,,\\
  \overline A_{i0}(t) &= \frac{a}{2}+\frac{b}{2\pi}\cos(\omega t)\,,\\
  \overline A_{ii}(t) &= \frac{1}{2}\,.
\end{align}
\end{subequations}
In principle, the
generalized replication coefficients
$\overline A_{\nu_1, \nu_2,\dots \nu_{k-1},j}(t)$ can be calculated to
arbitrary order for the landscape given in
Eq.~(\ref{eq:simple-single-peak}). However, the third order tensor has
already 8 independent entries, and with every higher order, the
number of independent entries doubles.

\end{appendix}


\begin{thebibliography}{10}

\bibitem{Eigen71}
M. Eigen, Naturwissenschaften {\bf 58},  465  (1971).

\bibitem{ThompsonMcBride74}
C.~J. Thompson and J.~L. McBride, {\frenchspacing Math. Biosci.} {\bf 21},  127
   (1974).

\bibitem{JonesEnnsRangnekar76}
B.~L. Jones, R.~H. Enns, and S.~S. Rangnekar, {\frenchspacing Bull. Math.
  Biol.} {\bf 38},  15  (1976).

\bibitem{Jones79a}
B.~L. Jones, {\frenchspacing Bull. Math. Biol.} {\bf 41},  761  (1979).

\bibitem{Jones79b}
B.~L. Jones, {\frenchspacing Bull. Math. Biol.} {\bf 41},  849  (1979).

\bibitem{EigenSchuster79}
M. Eigen and P. Schuster, {\em The Hypercycle---A Principle of Natural
  Self-Organization} (Springer-Verlag, Berlin, 1979).

\bibitem{SwetinaSchuster82}
J. Swetina and P. Schuster, {\frenchspacing Biophys. Chem.} {\bf 16},  329
  (1982).

\bibitem{McCaskill84}
J.~S. McCaskill, {\frenchspacing J. Chem. Phys.} {\bf 80},  5194  (1984).

\bibitem{Rumschitzki87}
D.~S. Rumschitzki, {\frenchspacing J. Math. Biol.} {\bf 24},  667  (1987).

\bibitem{Leuthaeusser87}
I. Leuth\"ausser, {\frenchspacing J. Stat. Phys.} {\bf 48},  343  (1987).

\bibitem{Eigenetal88}
M. Eigen, J. McCaskill, and P. Schuster, {\frenchspacing J. Phys. Chem.} {\bf
  92},  6881  (1988).

\bibitem{SchusterSwetina88}
P. Schuster and J. Swetina, {\frenchspacing Bull. Math. Biol.} {\bf 50},  635
  (1988).

\bibitem{Eigenetal89}
M. Eigen, J. McCaskill, and P. Schuster, {\frenchspacing Adv. Chem. Phys.} {\bf
  75},  149  (1989).

\bibitem{NowakSchuster89}
M. Nowak and P. Schuster, {\frenchspacing J. theor. Biol.} {\bf 137},  375
  (1989).

\bibitem{FranzPeliti97}
S. Franz and L. Peliti, {\frenchspacing J. Phys. A} {\bf 30},  4481  (1997).

\bibitem{AlvesFontanari98}
D. Alves and J.~F. Fontanari, {\frenchspacing Phys. Rev. E} {\bf 57},  7008
  (1998).

\bibitem{WilkeRonnewinkelMartinetz99}
C.~O. Wilke, C. Ronnewinkel, and T. Martinetz,  in {\em Advances in Artificial
  Life, Proceedings of ECAL'99, Lausanne, Switzerland}, {\em Lecture Notes in
  Artificial Intelligence}, edited by D. Floreano, J.-D. Nicoud, and F. Mondada
  (Springer-Verlag, New York, 1999), pp.\ 417--421.

\bibitem{NilssonSnoad2000}
M. Nilsson and N. Snoad, {\frenchspacing Phys. Rev. Lett.} {\bf 84},  191
  (2000).

\bibitem{BaakeGabriel99}
E. Baake and W. Gabriel, {\frenchspacing Ann. Rev. Comp. Phys.} {\bf 7},
  (1999), in press.

\bibitem{Domingoetal78}
E. Domingo, D. Sabo, T. Taniguchi, and C. Weissmann, Cell {\bf 13},  735
  (1978).

\bibitem{AdamiBrown94}
C. Adami and C.~T. Brown,  in {\em Artificial Life IV}, edited by R.~A. Brooks
  and P. Maes (MIT Press, Cambridge, MA, 1994), pp.\ 372--381.

\bibitem{Gomezetal99}
J. Gomez {\it et~al.}, {\frenchspacing J. of Viral Hepatitis} {\bf 6},  3
  (1999).

\bibitem{RonnewinkelWilkeMartinetz2000}
C. Ronnewinkel, C.~O. Wilke, and T. Martinetz,  in {\em Theoretical Aspects of
  Evolutionary Computing}, edited by L. Kallel, B. Naudts, and A. Rogers
  (Springer-Verlag, New York, 2000).

\bibitem{NilssonSnoad2000a}
M. Nilsson and N. Snoad, Quasispecies evolution on a fitness landscape with a
  fluctuating peak, eprint physics/0004039, April 2000.

\bibitem{NilssonSnoad2000b}
M. Nilsson and N. Snoad, Optimal Mutation Rates in Dynamic Environments, eprint
  physics/0004042, April 2000.

\bibitem{Wilke99}
C.~O. Wilke, {\em Evolutionary Dynamics in Time-Dependent Environments} (Shaker
  Verlag, Aachen, 1999), {PhD} thesis Ruhr-Universit\"at Bochum.

\bibitem{Hirst97a}
T. Hirst,  in {\em Fourth European Conference on Artificial Life}, edited by P.
  Husband and I. Harvey (MIT Press, ADDRESS, 1997), pp.\ 425--431.

\bibitem{HirstRowe99}
A.~J. Hirst and J.~E. Rowe, {\frenchspacing J. theor. Biol.}  (1998),
  submitted.

\bibitem{Schmittetal98}
L. Schmitt, C.~L. Nehaniv, and R.~H. Fujii, \frenchspacing Theor. Comp. Sci.
  {\bf 200},  101  (1998).

\bibitem{SchmittNehaniv99}
L. Schmitt and C.~L. Nehaniv,  in {\em Mathematical \& Computational Biology:
  Computational Morphogenesis, Hierarchical Complexity, and Digital Evolution},
  {\em Lectures on Mathematics in the Life Sciences}, edited by C.~L. Nehaniv
  (American Mathematical Society, ADDRESS, 1999), pp.\ 147--166.

\bibitem{Rowe99a}
J.~E. Rowe,  in {\em Proceedings of GECCO 1999}, edited by W. Banzhaf {\it
  et~al.} (Morgan Kaufmann, San Mateo, 1999), p.\ 557.

\bibitem{Rowe99b}
J.~E. Rowe,  in {\em Proceedings of the 2nd Evonet Summerschool}
  (Springer-Verlag, New York, 1999), in press.

\bibitem{Fontanaetal93}
W. Fontana {\it et~al.}, {\frenchspacing Phys. Rev. E} {\bf 47},  2083  (1993).

\bibitem{FontanaSchuster98}
W. Fontana and P. Schuster, Nature {\bf 280},  1451  (1998).

\bibitem{Forst98}
C.~V. {Forst}, J.\ Biotechnology {\bf 64},  101  (1998).

\bibitem{ForstReidysWeber95}
C.~V. Forst, C. Reidys, and J. Weber,  in {\em Advances in Artificial Life},
  Vol.~929 of {\em Lecture Notes in Artificial Intelligence}, edited by F.
  Mor{\'a}n, A. Moreno, J.~J. Merelo, and P. Chac{\'o}n (Springer, ADDRESS,
  1995), pp.\ 128--147, {S}FI Preprint 95-10-094.

\bibitem{HuynenStadlerFontana96}
M.~A. Huynen, P.~F. Stadler, and W. Fontana, {\frenchspacing Proc. Natl. Acad.
  Sci. USA} {\bf 93},  397  (1996).

\bibitem{ReidysStadlerSchuster97}
C. Reidys, P.~F. Stadler, and P. Schuster, {\frenchspacing Bull. Math. Biol.}
  {\bf 59},  339  (1997), {S}FI Preprint 95-07-058.

\bibitem{Nimwegenetal99}
E. van Nimwegen, J.~P. Crutchfield, and M. Huynen, {\frenchspacing Proc. Natl.
  Acad. Sci. USA} {\bf 96},  9716  (1999).

\bibitem{WilkeMartinetz99a}
C.~O. Wilke and T. Martinetz, {\frenchspacing Phys. Rev. E} {\bf 60},  2154
  (1999).

\bibitem{Erugin66}
N.~P. Erugin, {\em Linear System of Ordinary Differential Equations, with
  Periodic and Quasi-Periodic Coefficients} (Academic Press, New York, London,
  1966).

\bibitem{Perron07}
O. Perron, {\frenchspacing Math. Ann.} {\bf 64},  248  (1907).

\bibitem{DressRumschitzki88}
A.~W.~M. {Dress} and D.~S. {Rumschitzki}, Acta Applicandae Mathematicae {\bf
  11},  103  (1988).

\bibitem{Galluccio97}
S. Galluccio, {\frenchspacing Phys. Rev. E} {\bf 56},  4526  (1997).

\bibitem{MaynardSmith83}
J. {Maynard Smith}, {\frenchspacing Proc. R. Soc. London~B} {\bf 219},  315
  (1983).

\bibitem{Higgs94}
P.~G. Higgs, {\frenchspacing Genet. Res. Camb.} {\bf 63},  63  (1994).

\bibitem{WoodcockHiggs96}
G. Woodcock and P.~G. Higgs, {\frenchspacing J. theor. Biol.} {\bf 179},  61
  (1996).

\bibitem{WagnerBaakeGerisch97}
H. Wagner, E. Baake, and T. Gerisch, {\frenchspacing J. Stat. Phys.} {\bf 92},
  1017  (1998).

\bibitem{Leuthaeusser85}
I. Leuth\"ausser, {\frenchspacing J. Chem. Phys.} {\bf 84},  1884  (1985).

\bibitem{Tarazona92}
P. Tarazona, {\frenchspacing Phys. Rev. E} {\bf 45},  6038  (1992).

\bibitem{Baakeetal97}
E. Baake, M. Baake, and H. {Wagner}, {\frenchspacing Phys. Rev. Lett.} {\bf
  78},  559  (1997).

\bibitem{YakubovichStarzhinskii75}
Y.~A. Yakubovich and V.~M. Starzhinskii, {\em Linear Differential Equations
  with Periodic Coefficients} (John Wiley \& Sons, New York, 1975), Vol.~1.

\bibitem{Demetriusetal85}
L. Demetrius, P. Schuster, and K. Sigmund, {\frenchspacing Bull. Math. Biol.}
  {\bf 47},  239  (1985).

\bibitem{SasakiIwasa87}
A. Sasaki and Y. Iwasa, Genetics {\bf 115},  377  (1987).

\bibitem{IshiiMatsudaIwasaSasaki89}
K. Ishii, H. Matsuda, Y. Iwasa, and A. Sasaki, Genetics {\bf 121},  163
  (1989).

\bibitem{Charlesworth93}
B. Charlesworth, {\frenchspacing Genet. Res. Camb.} {\bf 61},  205  (1993).

\bibitem{LandeShannon96}
R. Lande and S. Shannon, Evolution {\bf 50},  434  (1996).

\bibitem{LevitanKauffman95}
B. Levitan and S. Kauffman, Molecular Diversity {\bf 1},  53  (1995).

\bibitem{Levitan97}
B. Levitan,  in {\em Annual Reports in Combinatorial Chemistry and Molecular
  Diversity}, edited by M.~R. Pavia, W.~H. Moos, A.~D. Ellington, and B.~K. Kay
  (ESCOM Publishers, The Netherlands, 1997), Vol.~1, pp.\ 95--152.

\bibitem{KauffmanLevin87}
S.~A. {Kauffman} and S. Levin, {\frenchspacing J. theor. Biol.} {\bf 128},  11
  (1987).

\bibitem{Kauffman92}
S.~A. Kauffman, {\em The Origins of Order} (Oxford University Press, Oxford,
  1992).

\bibitem{Gillespie91}
J.~H. Gillespie, {\em The Causes of Molecular Evolution} (Oxford University
  Press, Oxford, UK, 1991).

\bibitem{NimwegenCrutchfield99}
E. van Nimwegen and J.~P. Crutchfield, Metastable Evolutionary Dynamics:
  {C}rossing Fitness Barriers or Escaping via Neutral Paths?, eprint
  \emph{adap-org}/9907002, 1999.

\bibitem{Nimwegenetal97a}
E. van Nimwegen, J.~P. Crutchfield, and M. Mitchell, Theoretical Computer
  Science  (1997), to appear, SFI working paper 97-04-035.

\bibitem{Muller64}
H.~J. Muller, \frenchspacing{Mutat. Res.} {\bf 1},  2  (1964).

\bibitem{Wieheetal95}
T. Wiehe, E. Baake, and P. Schuster, {\frenchspacing J. theor. Biol.} {\bf
  177},  1  (1995).

\bibitem{Ronnewinkel99}
C. Ronnewinkel, unpublished, 1999.

\bibitem{Ewens79}
W.~J. Ewens, {\em Mathematical Population Genetics} (Springer-Verlag, New York,
  1979).

\bibitem{DerridaPeliti91}
B. Derrida and L. Peliti, {\frenchspacing Bull. Math. Biol.} {\bf 53},  355
  (1991).

\end{thebibliography}
\end{document}